\documentclass[twocolumn]{aastex631}

\graphicspath{{./}}

\usepackage{url}
\usepackage{natbib}
\usepackage{amsmath}
\usepackage{booktabs}
\usepackage{graphicx}
\usepackage{subfigure}
\usepackage{capt-of}

\begin{document}

\title{SN 2019tua : A Type IIb Supernova with Multiple Bumps in the Light Curves }

\author{Xin-Bo Huang}
\affiliation{Guangxi Key Laboratory for Relativistic Astrophysics, School of Physical Science and Technology, Guangxi University, Nanning, 530004, People’s Republic of China}
\affiliation{GXU-NAOC Center for Astrophysics and Space Sciences, Nanning, 530004, People’s Republic of China}

\author{Xiang-Gao Wang}
\affiliation{Guangxi Key Laboratory for Relativistic Astrophysics, School of Physical Science and Technology, Guangxi University, Nanning, 530004, People’s Republic of China}
\affiliation{GXU-NAOC Center for Astrophysics and Space Sciences, Nanning, 530004, People’s Republic of China}

\author{Long Li}
\affiliation{Department of Astronomy, University of Science and Technology of China, Hefei, 230026, People’s Republic of China}

\author{Li-Ping Xin}
\affiliation{CAS Key Laboratory of Space Astronomy and Technology, National Astronomical Observatories, Chinese Academy of Sciences, Beijing, 100101, People’s Republic of China}

\author{Jing Wang}
\affiliation{Guangxi Key Laboratory for Relativistic Astrophysics, School of Physical Science and Technology, Guangxi University, Nanning, 530004, People’s Republic of China}
\affiliation{CAS Key Laboratory of Space Astronomy and Technology, National Astronomical Observatories, Chinese Academy of Sciences, Beijing, 100101, People’s Republic of China}

\author{Tian-Ci Zheng}
\affiliation{Guangxi Key Laboratory for Relativistic Astrophysics, School of Physical Science and Technology, Guangxi University, Nanning, 530004, People’s Republic of China}
\affiliation{GXU-NAOC Center for Astrophysics and Space Sciences, Nanning, 530004, People’s Republic of China}

\author{Qi Wang}
\affiliation{Guangxi Key Laboratory for Relativistic Astrophysics, School of Physical Science and Technology, Guangxi University, Nanning, 530004, People’s Republic of China}
\affiliation{GXU-NAOC Center for Astrophysics and Space Sciences, Nanning, 530004, People’s Republic of China}

\author{Hui-Ya Liu}
\affiliation{Guangxi Key Laboratory for Relativistic Astrophysics, School of Physical Science and Technology, Guangxi University, Nanning, 530004, People’s Republic of China}
\affiliation{GXU-NAOC Center for Astrophysics and Space Sciences, Nanning, 530004, People’s Republic of China}

\author{Zi-Min Zhou}
\affiliation{Guangxi Key Laboratory for Relativistic Astrophysics, School of Physical Science and Technology, Guangxi University, Nanning, 530004, People’s Republic of China}
\affiliation{GXU-NAOC Center for Astrophysics and Space Sciences, Nanning, 530004, People’s Republic of China}

\author{Xiao-meng Lu}
\affiliation{CAS Key Laboratory of Space Astronomy and Technology, National Astronomical Observatories, Chinese Academy of Sciences, Beijing, 100101, People’s Republic of China}

\author{jian-yan Wei}
\affiliation{CAS Key Laboratory of Space Astronomy and Technology, National Astronomical Observatories, Chinese Academy of Sciences, Beijing, 100101, People’s Republic of China}

\author{En-Wei Liang}
\affiliation{Guangxi Key Laboratory for Relativistic Astrophysics, School of Physical Science and Technology, Guangxi University, Nanning, 530004, People’s Republic of China}
\affiliation{GXU-NAOC Center for Astrophysics and Space Sciences, Nanning, 530004, People’s Republic of China}
   
\correspondingauthor{Xiang-Gao Wang}
\email{wangxg@gxu.edu.cn}

\begin{abstract}

We present photometric and spectroscopic observations and analysis of the type IIb supernova (SN) SN 2019tua, which exhibits multiple bumps in its declining light curves between 40 and 65 days after discovery. 
SN 2019tua shows a time to peak of about 25 days similar to other type IIb SNe. 
Our observations indicate a decrease in its brightness of about 1 magnitude in the 60 days after the peak. 
At about days 50, and 60, its multiband light curves exhibit bumpy behavior. 
The complex luminosity evolution of SN 2019tua could not be well modeled with a single currently popular energy source model, e.g., radioactive decay of $^{56}$Ni, magnetar, interaction between the ejecta and a circumstellar shell. 
Even though the magnetar model has a smaller \( \chi^2 / \text{dof} \) value, the complex changes in SN 2019tua's brightness suggest that more than one physical process might be involved. 
We propose a hybrid CSM interaction plus $^{56}$Ni model to explain the bolometric light curve (LC) of SN 2019tua. 
The fitting results show that the ejecta mass $M_{\rm ej} \approx 2.4~M_\odot$, the total CSM mass $M_{\rm CSM} \approx 1.0~M_\odot$, and the $^{56}$Ni mass $M_{\rm Ni} \approx 0.4~M_\odot$. 
The total kinetic energy of the ejecta is $E_k\approx 0.5 \times 10^{51}\rm~erg$. 
Pre-existing multiple shells suggest that the progenitor of SN 2019tua experienced mass ejections within approximately $\sim6 - 44$ years prior to the explosion.

\end{abstract}


\section{Introduction} \label{sec:intro}
Supernovae (SNe), among the most luminous events in stellar evolution, are classified into distinct types based on their spectral characteristics and underlying explosion mechanisms. A notable subclass is type IIb SNe, which initially exhibit hydrogen lines in their spectra, later transitioning to helium-dominated features. This spectral evolution reflects the unique nature of their progenitors, stars that have retained just enough hydrogen to display its spectral lines post-explosion, but with hydrogen envelopes so depleted that helium lines eventually prevail. Type IIb SNe, thus, occupy a critical position in the study of massive star evolution and SN physics, offering insights into the processes driving the loss of stellar outer layers (see e.g., \citealt{1997ARA&A..35..309F,1993Natur.364..509P}). 
  
The light curves (LCs) of type IIb SNe are governed by the complex interplay of nuclear decay and dynamical processes. The predominant energy source in the later stages is the radioactive decay of $^{56}$Ni to $^{56}$Co, and ultimately to $^{56}$Fe (e.g., \citealt{1994ApJ...429..300W,1995PhR...256..173N}). 
Some type IIb SNe exhibit double-peaked LCs, characterized by a rapidly rising and falling peak before the main peak driven by $^{56}$Ni. The early emission is attributed to shock-cooling emission from the heated extended envelope of the progenitor, and is observed in some cases such as SN 1993J \citep{1994AJ....107.1022R,1994ApJ...429..300W}, SN 2011dh \citep{2011ApJ...742L..18A,2012ApJ...752...78S,2014A&A...562A..17E}, SN 2013df \citep{2014MNRAS.445.1647M,2014AJ....147...37V}, SN 2016gkg \citep{2022ApJ...934..186N,2022ApJ...936..111K}, SN 2017jgh \citep{2021MNRAS.507.3125A}, and ZTF18aalrxas \citep{2019ApJ...878L...5F}. 
Additionally, interactions with dense CSM around the progenitor star introduce complex shock structures (both forward and reverse), transforming kinetic energy into thermal radiation and significantly influencing the LC (e.g., \citealt{1994ApJ...420..268C,2011MNRAS.415..199M}). Energy diffusion from the densely packed ejecta also plays a role, especially in the early to mid-phases of the LC. These energy contributions, varying in their relative impact based on SN properties such as \( ^{56}\text{Ni} \) mass, CSM density, and explosion energy, collectively shape the unique LC signature of type IIb SNe.

Type IIb SNe progenitors are massive stars that have undergone substantial hydrogen envelope loss, resulting in a residual thin hydrogen layer at the explosion. This phenomenon occurs in two primary contexts: single massive stars, often exceeding 20 $M_{\odot}$, where strong stellar winds or eruptions induced by nuclear burning phases precipitate significant hydrogen loss \citep{2003ApJ...591..288H,2011ApJ...732...63S}, and binary systems where a massive star transfers much of its hydrogen envelope to a companion, a process considered more prevalent and efficient for hydrogen stripping \citep{1992ApJ...391..246P,2013MNRAS.436..774E}. This binary interaction paradigm, responsible for many type IIb SNe, aligns with observations and theoretical models \citep{2004Natur.427..129M,2015MNRAS.447..772F}, highlighting the diverse evolutionary pathways leading to these SNe.
High-resolution imaging, notably from the Hubble Space Telescope, has identified diverse progenitors for SNe IIb. Studies have revealed supergiants in binary systems for SNe IIb 1993J \citep{2009Sci...324..486M} and 2001ig \citep{2018ApJ...856...83R}, a stripped, Wolf-Rayet-like star for SN IIb 2008ax \citep{2008MNRAS.391L...5C,2008MNRAS.389..955P}, and yellow supergiants for SNe IIb 2011dh \citep{2011ApJ...739L..37M,2014ApJ...793L..22F}, 2013df \citep{2015ApJ...807...35M}, and 2016gkg \citep{2022ApJ...936..111K,2022ApJ...934..186N}, indicating a range of initial masses and binary interactions.

Bumpy LCs, marked by noticeable fluctuations in brightness over time, are a relatively common feature in certain types of SNe, particularly in superluminous SNe and type IIn SNe. 
Recent research has shown that the LCs of hydrogen-poor superluminous SNe, which were thought to be powered by magnetar central engines, do not always follow the smooth decline predicted by a simple magnetar spin-down model \citep{2022ApJ...933L..45H}. 
\cite{2022ApJ...933L..45H} systematic study of 34 superluminous supernovae (SLSNe) revealed that the majority of these events cannot be explained solely by the smooth magnetar model, and they exhibit irregular bumps in their post-peak LCs. The cause of these bumps could be intrinsic to the SN (e.g., a variable central engine) or extrinsic (e.g., circumstellar interaction). Some SNe IIn, such as SN 2005la \citep{2008MNRAS.389..131P}, SN 2006jd \citep{2012ApJ...756..173S}, SN 2009ip \citep{2014ApJ...797..107M,2015AJ....149....9M}, and iPTF13z \citep{2017A&A...605A...6N}, display varied re-brightening episodes after peak brightness, ranging from short, early bumps to long, late ones. These irregularities in the LC are thought to arise from complex interactions between the SN ejecta and the circumstellar material. However, in the context of type IIb SNe, the occurrence of bumpy LCs is notably rare, with no well-documented cases in the existing IIb SN sample.

In this paper, we present follow-up observations and studies of the type IIb SN SN 2019tua, which exhibits multiple bumps in its LCs.
In Section \ref{sec:observations}, we summarize the observations of SN 2019tua.
In Section \ref{sec:analysis}, we analyze the observations and the properties of LCs and spectra, and includes a blackbody fitting of the spectral energy distribution.
In Section \ref{sec:Modelling of SN 2019tua}, We described some simple physical LC models, and then used these models to fit the bolometric LC and relevant parameters.
Our discussion and conclusions can be found in Section \ref{sec:discussion}.

\section{Observations and Data Reduction} \label{sec:observations}
\subsection{Optical Photometry}

\begin{figure}
    \centering
    
    \subfigure[The GWAC F60B image in \textit{B} band.\label{fig:sub1}]{\fbox{\includegraphics[width=0.4\linewidth]{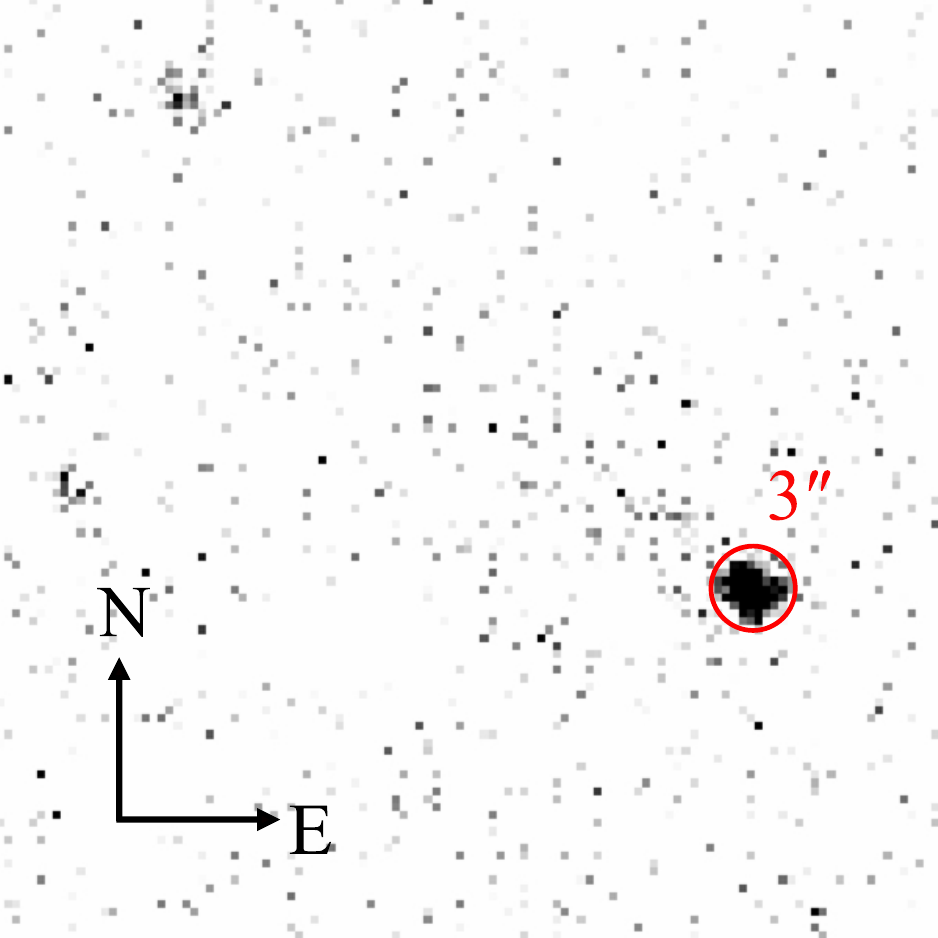}}}
    \hspace{1em}
    \subfigure[The GWAC F60B image in \textit{V} band.\label{fig:sub2}]{\fbox{\includegraphics[width=0.4\linewidth]{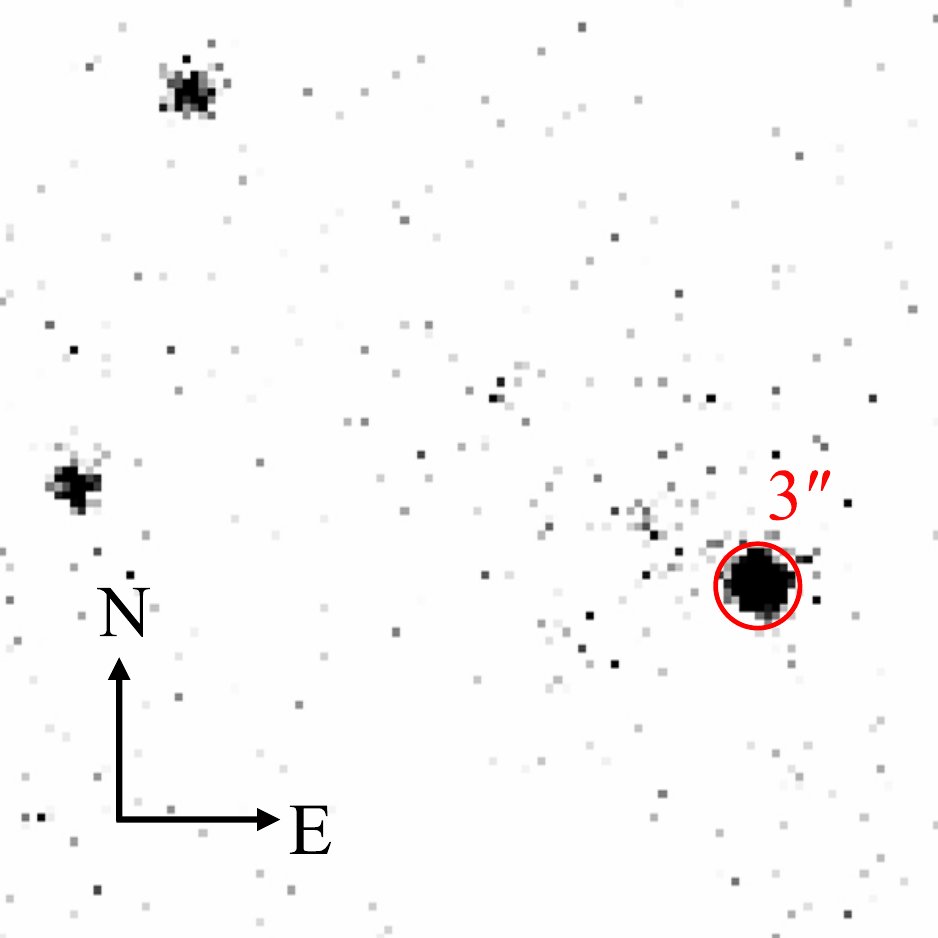}}}
    
    \vspace{1em}
    
    \subfigure[The GWAC F60B image in \textit{R} band.\label{fig:sub3}]{\fbox{\includegraphics[width=0.4\linewidth]{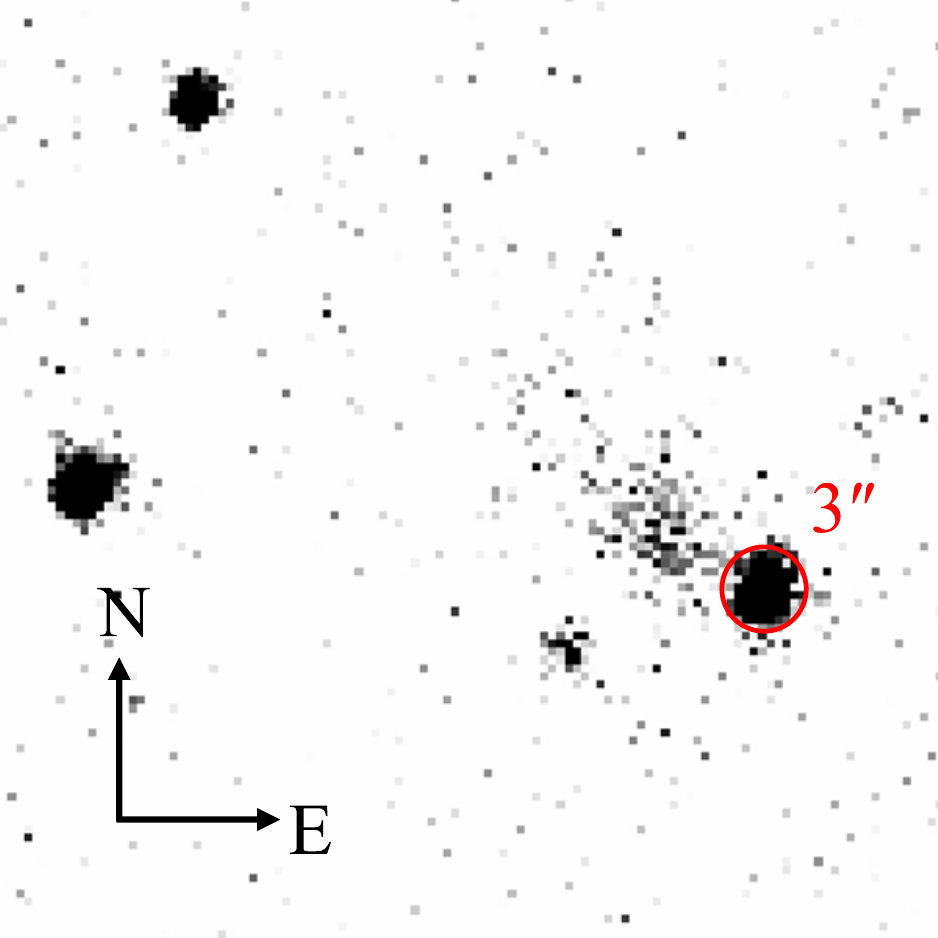}}}
    \hspace{1em}
    \subfigure[The GWAC F60B image in \textit{I} band.\label{fig:sub4}]{\fbox{\includegraphics[width=0.4\linewidth]{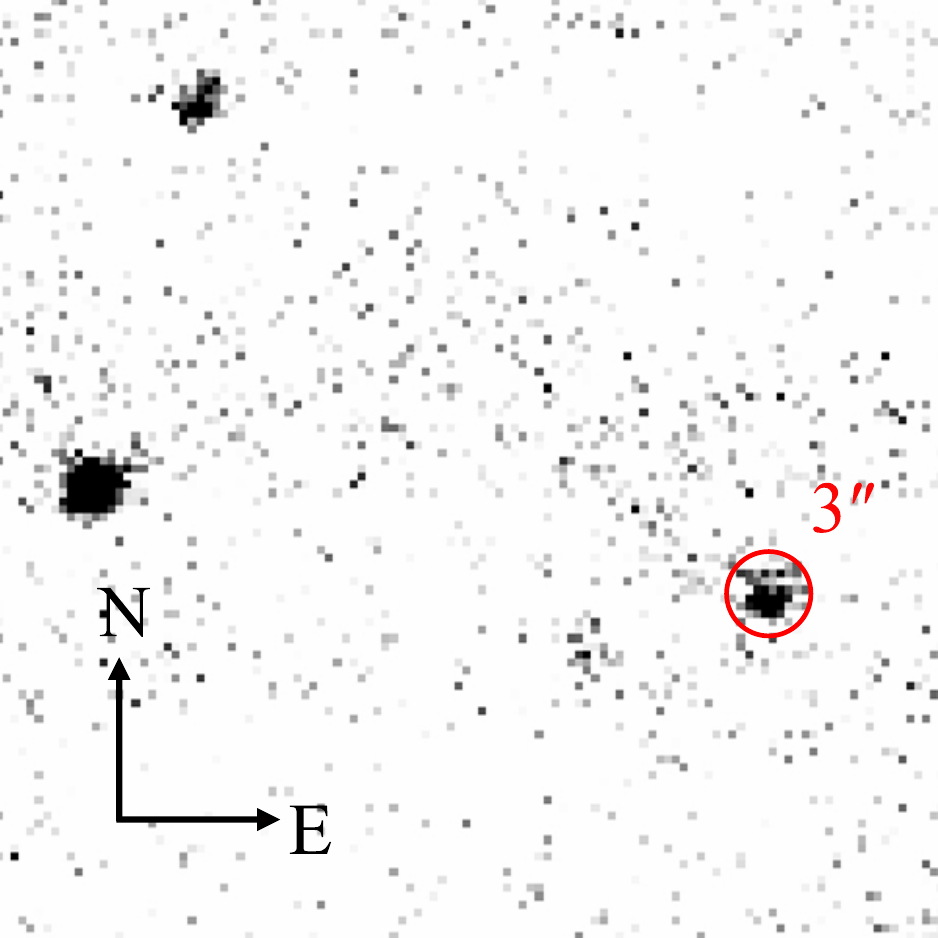}}}
    
    \caption{The \textit{BVRI} bands images of the SN 2019tua obtained by the GWAC F60B telescope during its first observation on November 3, 2019, at 14:16:41.60 UT. These \textit{BVRI} band images are obtained by combining three 150 s exposures each. The localization of the SN 2019tua is marked with a red circle in each panel, with a diameter of 3 arcseconds.}
    \label{fig:optical}
\end{figure}

\begin{figure*}[htp]
    \centering
    \includegraphics[width=18cm]{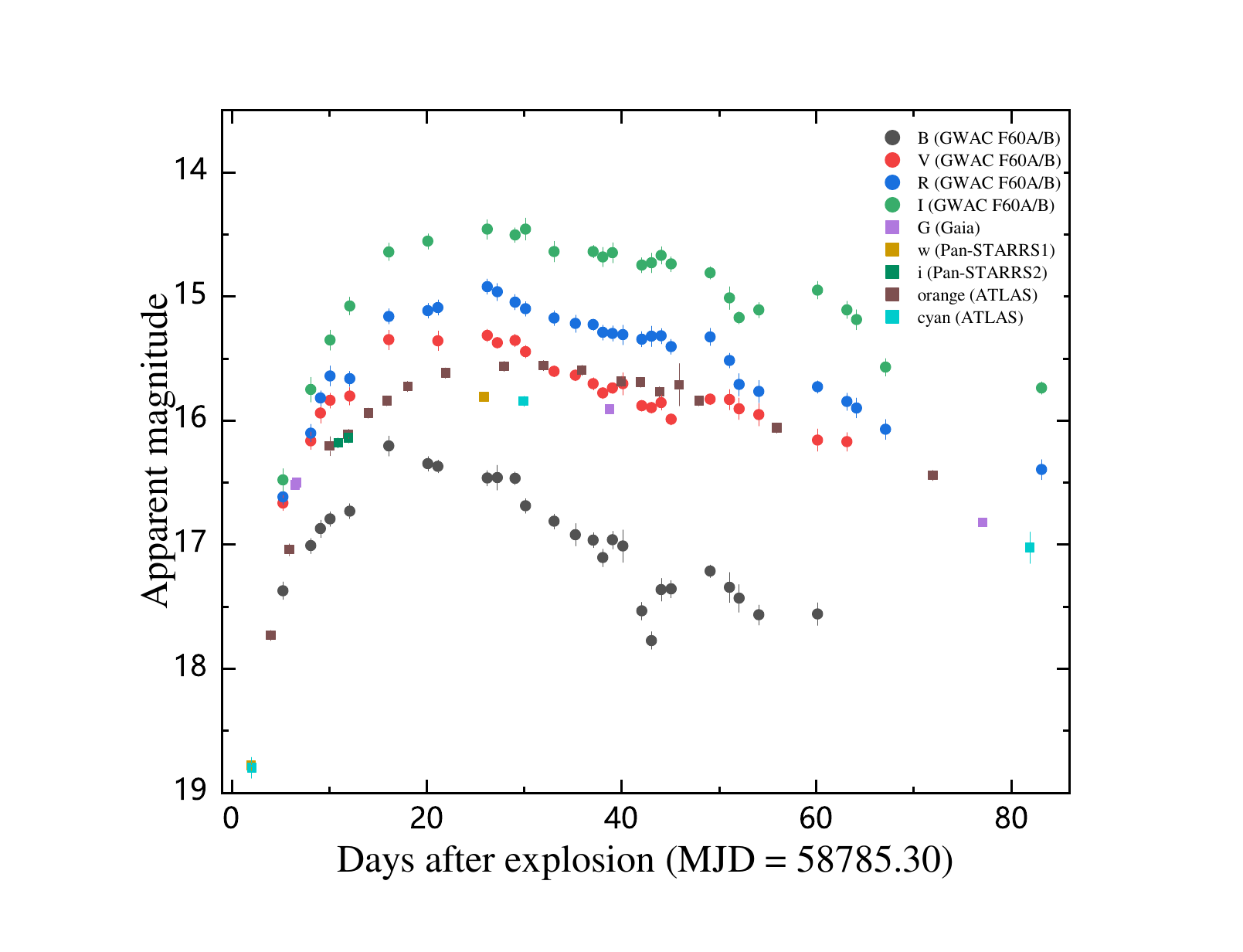}
    \caption{Multi-band LCs of SN 2019tua observed by GWAC F60A and F60B telescope are presented, with corrections for Galactic line-of-sight extinction. Our observations results are shown as circles, while squares markers represent data sourced from the TNS and some of other online report. Due to the dense data points from ATLAS, we binned and averaged the data taken within a single day, and then plotted these data.}
    \label{fig:lcs}
\end{figure*}
\par SN 2019tua was discovered on October 31, 2019, at 07:52:19 UT (MJD = 58787.33) by \citealt{2019TNSTR2232....1T}, using the Asteroid Terrestrial-impact Last Alert System (ATLAS; \citealt{2018PASP..130f4505T}) with the ATLAS-02 telescope at the Haleakala Observatory \citep{2021PASP..133b9201S}. 

\begin{deluxetable*}{cccccccc}[htp]
\tablecaption{Optical Photometry of SN 2019tua\label{tab:lightcure message}}
\tablehead{
\colhead{ UT Date } & \colhead{ Time} & \colhead{ $B$ } & \colhead{ $V $} &
\colhead{ $R$ } & \colhead{ $I$ } & \colhead{ Telescope }  \\
\colhead{  } & \colhead{(days)} & \colhead{(mag)} & \colhead{(mag)} &\colhead{(mag)} & \colhead{(mag)} 
& \colhead{}}
\startdata
 2019-11-03 & 5.26  & 17.37$\pm0.07$ & 16.66$\pm0.06$ & 16.62$\pm0.06$ & 16.48$\pm0.09$  & GWAC F60B \\
 2019-11-06 & 8.13  & 17.02$\pm0.08$ & 16.16$\pm0.07$ & 16.10$\pm0.07$ & 15.75$\pm0.10$  & GWAC F60B  \\
 2019-11-07 & 9.10  & 16.87$\pm0.07$ & 15.94$\pm0.08$ & 15.82$\pm0.06$ & 15.18$\pm0.09$  & GWAC F60B  \\
 2019-11-08 & 10.09 & 16.79$\pm0.06$ & 15.84$\pm0.06$ & 15.64$\pm0.08$ & 15.35$\pm0.08$  & GWAC F60B \\
 2019-11-10 & 12.12 & 16.73$\pm0.06$ & 15.80$\pm0.07$ & 15.66$\pm0.06$ & 15.08$\pm0.07$  & GWAC F60B  \\
 2019-11-14 & 16.14 & 16.21$\pm0.08$ & 15.35$\pm0.08$ & 15.16$\pm0.06$ & 14.64$\pm0.07$  & GWAC F60B  \\
 2019-11-18 & 20.13 & 16.35$\pm0.06$ & \nodata      & 15.12$\pm0.06$ & 14.56$\pm0.06$  & GWAC F60B  \\
 2019-11-19 & 21.14 & 16.37$\pm0.05$ & 15.36$\pm0.08$ & 15.09$\pm0.06$ & \nodata       & GWAC F60B  \\
 2019-11-24 & 26.22 & 16.46$\pm0.06$ & 15.32$\pm0.05$ & 14.92$\pm0.06$ & 14.46$\pm0.08$  & GWAC F60B  \\
 2019-11-25 & 27.26 & 16.46$\pm0.10$ & 15.37$\pm0.04$ & 14.96$\pm0.07$ & \nodata       & GWAC F60B   \\
 2019-11-27 & 29.08 & 16.46$\pm0.05$ & 15.36$\pm0.05$ & 15.05$\pm0.06$ & 14.51$\pm0.06$  & GWAC F60B   \\
 2019-11-28 & 30.17 & 16.69$\pm0.06$ & 15.45$\pm0.05$ & 15.10$\pm0.06$ & 14.46$\pm0.09$  & GWAC F60B   \\
 2019-12-01 & 33.09 & 16.81$\pm0.06$ & 15.60$\pm0.04$ & 15.18$\pm0.06$ & 14.64$\pm0.08$  & GWAC F60B   \\
 2019-12-03 & 35.27 & 16.92$\pm0.09$ & 15.64$\pm0.04$ & 15.22$\pm0.07$ & \nodata       & GWAC F60B \\
 2019-12-05 & 37.09 & 16.96$\pm0.06$ & 15.70$\pm0.05$ & 15.23$\pm0.03$ & 14.64$\pm0.05$  & GWAC F60B  \\
 2019-12-06 & 38.09 & 17.11$\pm0.07$ & 15.78$\pm0.04$ & 15.29$\pm0.06$ & 14.68$\pm0.08$  & GWAC F60B  \\
 2019-12-07 & 39.08 & 16.96$\pm0.07$ & 15.74$\pm0.04$ & 15.30$\pm0.06$ & 14.65$\pm0.08$  & GWAC F60B   \\
 2019-12-08 & 40.14 & 17.01$\pm0.13$ & 15.70$\pm0.09$ & 15.31$\pm0.08$ & \nodata       & GWAC F60B  \\
 2019-12-09 & 41.17 & 17.29$\pm0.07$ & 15.85$\pm0.05$ & \nodata      & 14.75$\pm0.04$  & GWAC F60B  \\
 2019-12-10 & 42.08 & 17.54$\pm0.07$ & 15.88$\pm0.04$ & 15.35$\pm0.06$ & 14.75$\pm0.06$  & GWAC F60B  \\
 2019-12-11 & 43.08 & 17.77$\pm0.07$ & 15.90$\pm0.04$ & 15.32$\pm0.08$ & 14.73$\pm0.08$  & GWAC F60B  \\
 2019-12-12 & 44.08 & 17.36$\pm0.09$ & 15.86$\pm0.06$ & 15.32$\pm0.06$ & 14.66$\pm0.03$  & GWAC F60B  \\
 2019-12-13 & 45.08 & 17.36$\pm0.07$ & 15.99$\pm0.04$ & 15.41$\pm0.06$ & 14.74$\pm0.04$  & GWAC F60B  \\
 2019-12-17 & 49.10 & 17.21$\pm0.05$ & 15.83$\pm0.04$ & 15.33$\pm0.07$ & 14.81$\pm0.0$5  & GWAC F60B  \\
 2019-12-19 & 51.08 & 17.34$\pm0.12$ & 15.83$\pm0.08$ & 15.52$\pm0.06$ & 15.01$\pm0.09$  & GWAC F60A \\
 2019-12-20 & 52.08 & 17.43$\pm0.11$ & 15.91$\pm0.08$ & 15.71$\pm0.09$ & 15.17$\pm0.05$  & GWAC F60A \\
 2019-12-22 & 54.09 & 17.57$\pm0.08$ & 15.95$\pm0.09$ & 15.77$\pm0.09$ & 15.11$\pm0.06$  & GWAC F60A  \\
 2019-12-28 & 60.10 & 17.56$\pm0.09$ & 16.16$\pm0.09$ & 15.73$\pm0.05$ & 14.95$\pm0.07$  & GWAC F60B  \\
 2019-12-31 & 63.12 & \nodata & 16.17$\pm0.07$ & 15.85$\pm0.07$ & 15.11$\pm0.07$ & GWAC F60A \\
 2020-01-01 & 64.11 & \nodata &      \nodata & 15.90$\pm0.08$ & 15.19$\pm0.08$ & GWAC F60A  \\
 2020-01-04 & 67.11 & \nodata &      \nodata & 16.07$\pm0.08$ & 15.57$\pm0.07$ & GWAC F60A \\
 2020-01-20 & 83.11 & \nodata &      \nodata & 16.40$\pm0.08$ & 15.74$\pm0.05$ & GWAC F60A \\
\enddata 
\tablecomments{The GWAC F60A and F60B telescope photometry in \textit{BVRI} bands. All photometry was corrected for the Galactic extinction. The time is relative to the reference time $T_0$ (MJD = 58785.30).}
\end{deluxetable*}

\par Its initial brightness, as filtered through the $\textit{cyan}$, registered at a magnitude of 18.63 from the forced photometry obtained from the ATLAS forced-photometry server \citep{2019TNSAN.119....1S}. 
It was first detected with $\textit{cyan} = 18.63\pm{0.139}$ at R.A. $21^{\mathrm{h}} 58^{\mathrm{m}} 00_{{\raisebox{0.35ex}{.}}}^{\mathrm{s}}280$ and Dec. $+24^{\circ} 15^{\prime} 57_{{\raisebox{0.35ex}{.}}}^{\prime \prime} 10$.
However, before the detection of SN 2019tua by ATLAS, the most recent observation of the same region by ATLAS was on October 29, 2019, at 07:09:07 (MJD = 58785.30), put a limit of brightness in the $\textit{orange}$ exceeding 19.49 \citep{2019TNSAN.119....1S}.
Thus, an explosion date ($T_0$) of 2019 October 29, at 07:09:07 UT (MJD = 58785.30) $\pm$ 2.0 days is adopted for the remainder of this work. The error listed on the explosion date is a flat distribution without any obvious bias toward the last nondetection or discovery.
And it was also monitored by the Panoramic Survey Telescope and Rapid Response System (Pan-STARRS; \citealt{2016arXiv161205560C}).
There is a first detection with Pan-STARRS on October 31, 2019, at 07:01:55 UT, 0.01 days before the first detection in $\textit{cyan}$, and put a brightness of \textit{w} = 18.78 mag.
\par The SN 2019tua was observed in the $\textit{BVRI}$ bands using the Ground-based Wide Angle Camera (GWAC) F60A and F60B telescope at the Xinglong Observatory of National Astronomical Observatories \citep{2020PASP..132e4502X}.
The \textit{BVRI} bands images obtained from the first observation with the F60B telescope are shown in Figure \ref{fig:optical}.  
Our optical observations began 5.26 days after $T_0$.
The \textit{B}-band observations of SN 2019tua extended for 55 days before the object became too faint for further observation, whereas the \textit{BVI} optical bands were monitored for a span of 80 days, and several images were taken during each night, with an individual exposure time of 150 s frame. 
We combined multiple 150 s exposures obtained from observations for subsequent photometric processing.
And the flat-field images were taken at dusk or dawn, while the bias images were taken at the beginning and the end of each observation.

\begin{figure*}[htp]
    \centering
    \includegraphics[width=16cm]{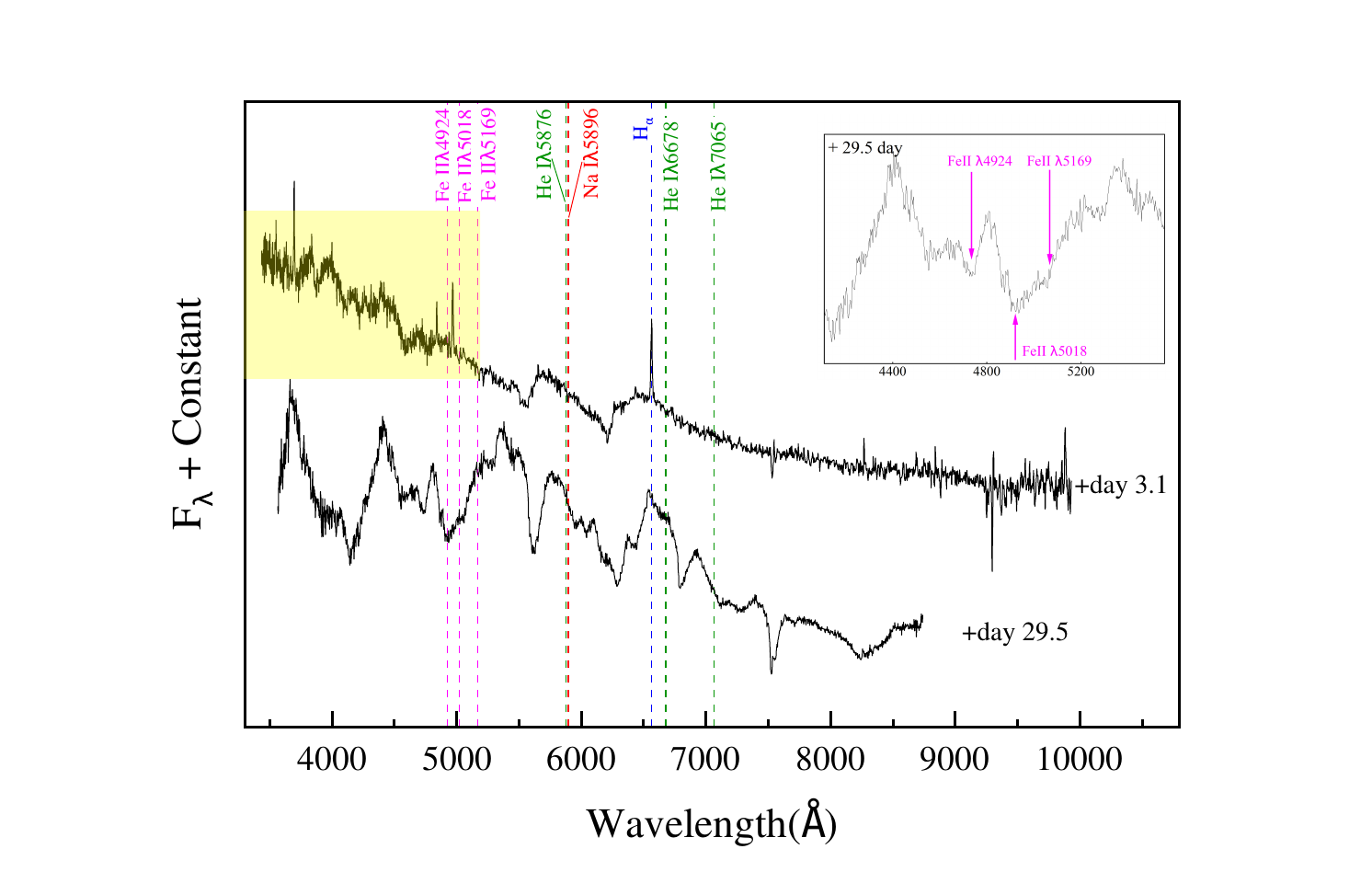}
    \caption{We obtained the spectrum of SN 2019tua using the Xinglong 2.16m telescope on day 29.5. For ease of analysis, we included the spectrum observed by the FTN telescope on day 3.3. And the lower spectrum in the figure is the spectrum observed by the Xinglong 2.16m telescope, while the upper one is the spectrum observed by the FTN telescope. The narrow emission lines within the yellow region may possibly originate from the host H II regions. The figure employs colored vertical lines to indicate the relevant Fe II, Na ID, He I, and H$\alpha$ lines. And the inset depicts the Fe II $\lambda$4924, Fe II $\lambda$5018 and Fe II $\lambda$5169 lines that we have identified in the day 29.5 spectrum, and they are marked with purple arrows consistent with the main figure.} 
    \label{fig:spec.pdf}
\end{figure*}

\begin{figure*}[htp]
    \centering
    \includegraphics[width=16cm]{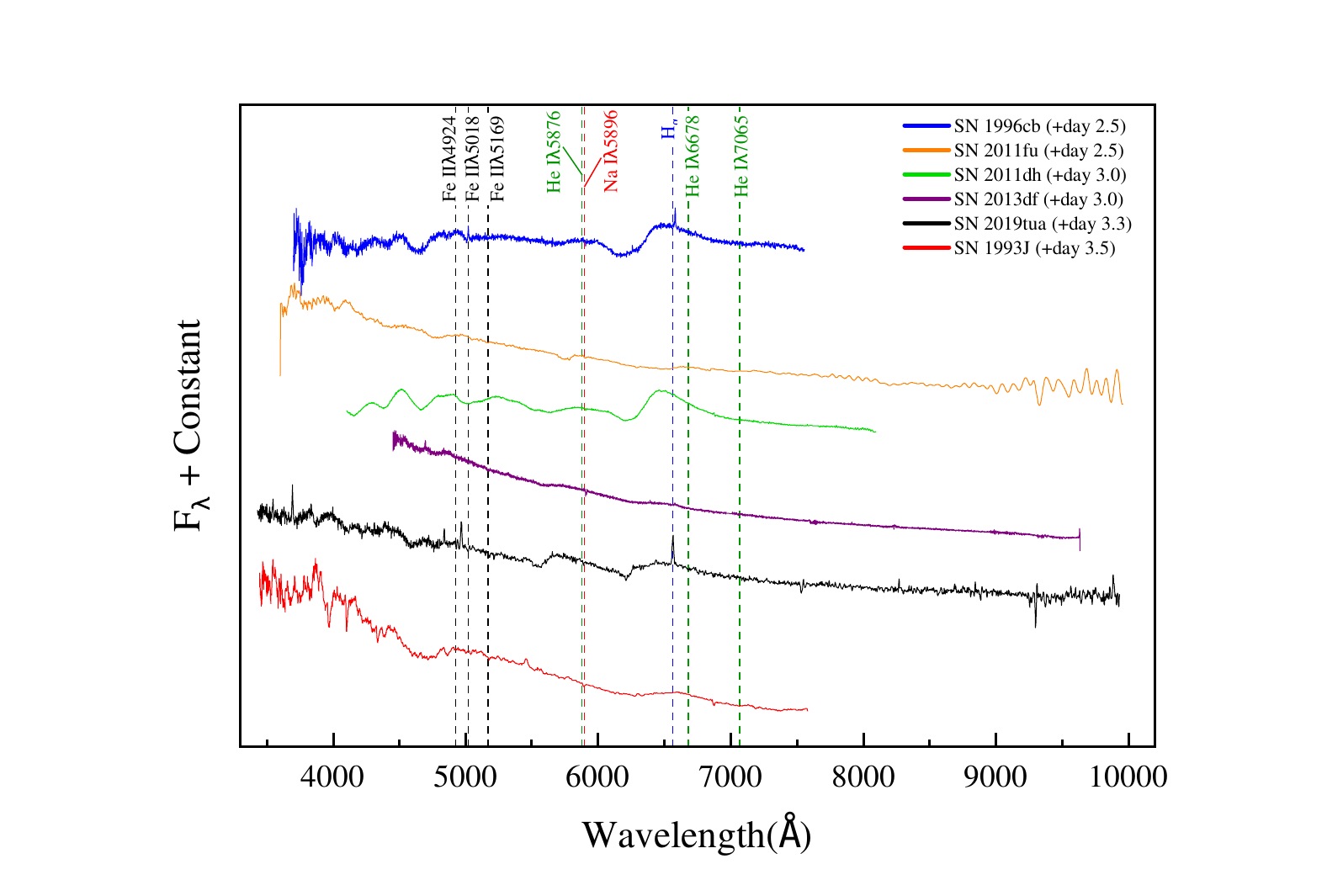}
    \caption{Early-time spectra of SN 2019tua compared with spectrum of other SNe IIb. Phase from explosion date for individual SNe is given in the legend, and notable spectral features are identified with color vertical lines at their rest-frame wavelengths.}
    \label{fig:spec1.pdf}
\end{figure*}

\begin{figure*}[htp]
    \centering
    \includegraphics[width=16cm]{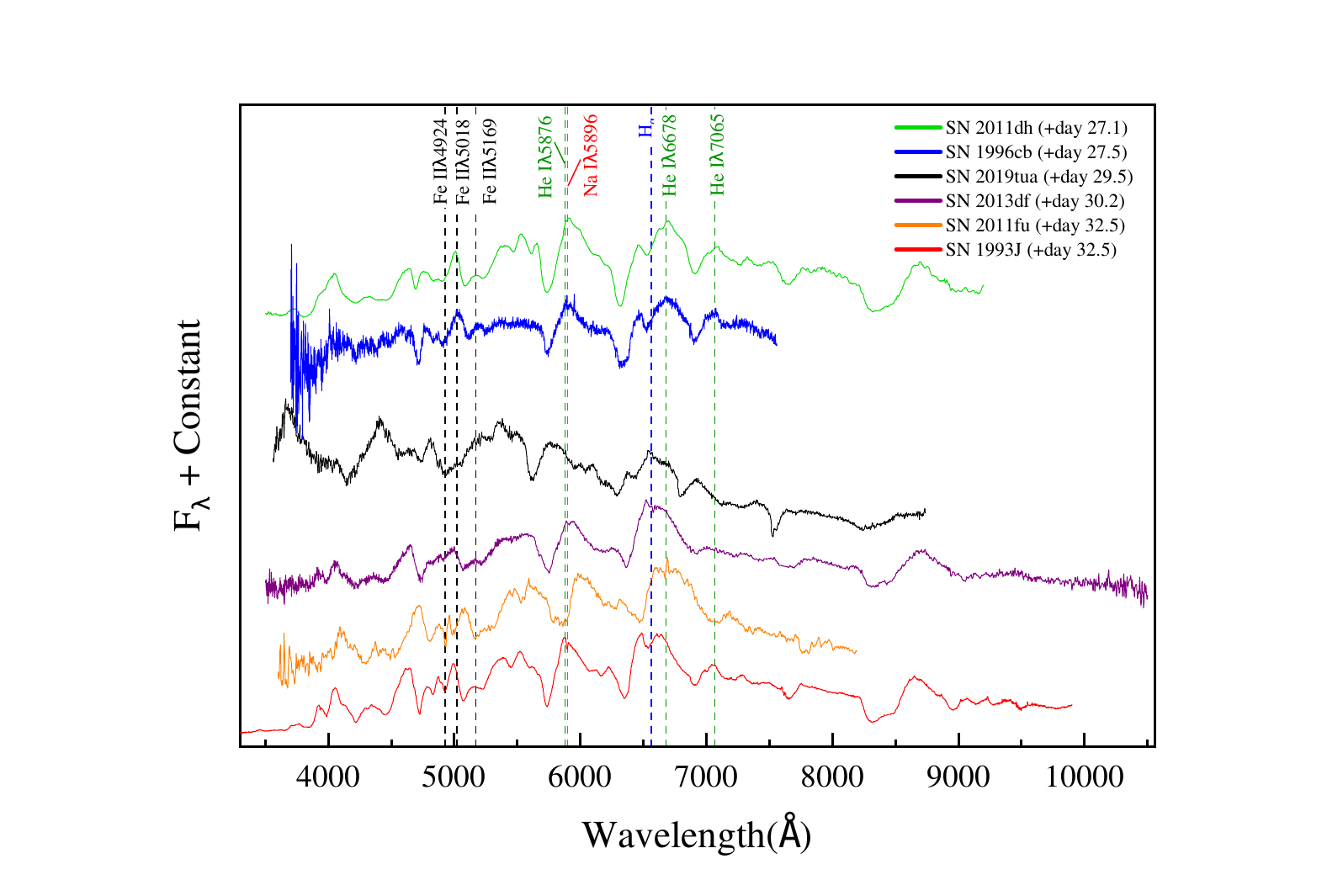}
    \caption{Comparison at about $\sim$ 3 days post-peak of the spectrum of SN 2019tua  with spectra of other SNe IIb. Phase from explosion date for individual SNe is given in the legend, and notable spectral features are identified with color vertical lines at their rest-frame wavelengths.}
    \label{fig:spec2.pdf}
\end{figure*}

\par Photometric data processing was performed the aperture photometry, with the \textit{apphot} task from the \textit{IRAF} \citep{1986SPIE..627..733T}.
Photometric calibration was performed using the USNO B1.0 \citep{2003AJ....125..984M} and NOMAD \citep{2004AAS...205.4815Z} catalogs and the results are reported in Table \ref{tab:lightcure message} and the processed optical LCs are shown in Figure \ref{fig:lcs}.
Moreover, these photometric data were corrected for Galactic extinction correction with ${E(B - V)_{\text{MW}}} = 0.06$ \citep{2011ApJ...737..103S}.
\par Additional observations of SN 2019tua were obtained in the \textit{cyan} and \textit{orange} filters in the ATLAS survey, the \textit{w} and \textit{i} filters with the Pan-STARRS system on the Pan-STARRS1 (PS1) telescope and Pan-STARRS2 (PS2) telescope, the \textit{G} filter on the Gaia's photometric instrument by the Gaia Science Alerts Project \citep{2019IAUS..339...12B}.
And the \textit{cyan}, \textit{orange}, \textit{w}, \textit{i} and \textit{G} forced-photometry light curves are shown in Figure \ref{fig:lcs}, respectively.

\subsection{Optiacl Spectroscopy}
\par We obtained an optical spectrum of SN 2019tua by using the Beijing Faint Object Spectrograph and Camera (BFOSC) on the Xinglong 2.16m telescope at the Xinglong Observatory of National Astronomical Observatories on November 27, 2019, at 18:39:43 UT (MJD = 58814.78) \citep{2016PASP..128k5005F}.
The exposure time is 30 min.
With a slit width of 1.8 arcseconds oriented in the north-south direction, the spectral resolution is approximately 10 Å when grating G4 was used, which results in a wavelength coverange of 3850 Å to 8000 Å.
Wavelength calibration was conducted by employing an iron-argon comparison lamp.
After bias subtraction and flat-field correction, we processed the two-dimensional spectrum via the \textit{IRAF} package \citep{1986SPIE..627..733T}. 
The isolated one-dimensional spectrum underwent wavelength and flux calibration through the corresponding comparison lamp and standard calibration star.
Moreover, the spectrum was corrected for the redshift correction with $z = 0.010$ \citep{2019TNSCR2244....1B} and the Galactic extinction correction with ${E(B - V)_{\text{MW}}} = 0.06$ \citep{2011ApJ...737..103S}. 
The resulting spectrum is shown in Figure \ref{fig:spec.pdf}.
\par We also included the spectrum that was uploaded to the Transient Name Serve (TNS \footnote{\url{www.wis-tns.org}}) by \citep{2019TNSCR2244....1B} in our analysis, and it was observed by using the Folded Low Order whYte-pupil Double-dispersed Spectrograph (FLOYDS) on the 2 m Faulkes Telescope North (FTN) at Haleakala with a wavelength range of 3500 Å to 10000 Å (\cite{2013PASP..125.1031B}).
And SN 2019tua exhibited a weak H$\alpha$ line in this early spectrum at day 3.3, which is consistent with typical type II Supernovae \citep{2019TNSCR2244....1B}.

\section{Analysis}\label{sec:analysis} 
\subsection{Light Curves Properties} \label{subsec:Optical multiband light curves} 
The initial observation and the final non-detection, combined with a limiting magnitude of 19.73 in the \textit{orange} band on the last day of non-detection \citep{2019TNSTR2232....1T}, argue against the likelihood of a pre-existing shock-cooling phase, such as those observed in SN 2008ax \citep{2008MNRAS.389..955P}, SN 2011dh \citep{2012ApJ...757...31B}, and SN 2020acat \citep{2022MNRAS.513.5540M}. 
Data from the \textit{BVRI} bands, first collected approximately 5 days subsequent to the estimated date of explosion, also exhibit no deviation from a rapid luminosity increase, also negating the presence of extended shock-cooling emissions.
\par In Figure \ref{fig:lcs}, SN 2019tua reached its optical peak around 25 days post-explosion, displaying a typical rise time for type IIb SN of $22.9 \pm 0.8$ days post-explosion \citep{2015A&A...574A..60T}.
After reaching the peak, it exhibited significant bumps in the LC at around 40 - 50 days and 50 - 65 days.
In this paper, the bumps we refer to are identified through visual inspection.
We designate as ``bumps'' the pronounced fluctuations in brightness occurring after the peak luminosity. Notably, there are conspicuous bumps both near and after the peak brightness of SN 2019tua.
Our analysis only focuses on interpreting the bumps around 50 days post-explosion.
The post-peak bump observed in SN 2019tua reminds the post-peak bump behavior seen in some of type IIn SNe, such as SN 2005la \citep{2008MNRAS.389..131P}, SN 2006jd \citep{2012ApJ...756..173S}, SN 2009ip \citep{2014ApJ...797..107M,2015AJ....149....9M}, and iPTF13z \citep{2017A&A...605A...6N}.
These bumpy behaviors are typically attributed to interactions between SN ejecta and CSM \citep{2018ApJ...856...59L,2019ApJ...885...43A,2023NatAs...7..779L}.
Due to the brightness variation of SN 2019tua around 50 days post-explosion, We will explain these two bumps during the 40-50 and 50-65 days period post-explosion through model fitting in the Section \ref{sec:Modelling of SN 2019tua}.

\subsection{Optical Spectral Properties} \label{subsec:optical spectrum}
Figure \ref{fig:spec.pdf} illustrates the spectral evolution of SN 2019tua from the rising phase ($t$ = 3.1 days) to the peaking phase that 3.3 days after the peaking day ($t$ = 29.5 days ). 
In the subsequent spectrum at day 29.5, the H$\alpha$ line weakened, while the presence of multiple He absorption features justified its classification as a type IIb supernova. Our classification (type IIb) is differ from the Transient Name Server (TNS) classification (type II) and we will further discuss to demonstrate that our modification to the original classification.
In order to further classify SN 2019tua, we compared its spectrum with the spectra of five similar-phase type IIb SNe (in Figure \ref{fig:spec1.pdf} and \ref{fig:spec2.pdf}).
These spectra were obtained from sne.space \footnote{\url{https://github.com/astrocatalogs/supernovae}}.
We conducted a search for well-documented and renowned type IIb SNe during periods close to the observation times of the two key spectra of SN 2019tua (2.5-5.0 days and 25-35 days).
The early spectrum of SN 2019tua at day 3.1 exhibits a prominent H$\alpha$ line, which is stronger compared to other IIb type SNe at similar phase, resembling the spectral features of SN 1996cb \citep{2019TNSCR2244....1B}. 
The presence of strong hydrogen lines also associates SN 2019tua with other typical type II SNe.
During this phase, in addition to the prominent, broad P-Cygni profile of the H$\alpha$ feature in its spectrum, the prominent broad feature of He I $\lambda$5876 is also evident, indicating a thin hydrogen envelope for SN 2019tua prior to the explosion \citep{2014ApJ...786L..15G,2018MNRAS.475.2344V}.
In the subsequent spectrum of SN 2019tua at 29.5 days, the H$\alpha$ line weakened and strong features of He I $\lambda$$\lambda$5876, 6678, 7065 were observed.
At this juncture, the He I $\lambda$6678 emission line is broader than H$\alpha$, and the H$\alpha$ P-Cygni absorption feature is deeper than the He I $\lambda$5876 absorption. 
This suggests that the He I $\lambda$6678 feature is discernible within the emission component of the H$\alpha$ P-Cygni profile, indicating that H and He exist in distinct shells within SN 2019tua \citep{2013MNRAS.433....2S,2022MNRAS.513.5540M}. 
Furthermore, the spectrum of SN 2019tua at day 29.5 revealed a blueshifted “notched” between H$\alpha$ and He I $\lambda$6678, which is also a useful signature for distinguishing SNe as type IIb SNe \citep{2017hsn..book..195G}.
This feature is also evident in the spectra of other type IIb SNe at similar epochs, as shown in Figure \ref{fig:spec2.pdf}.
In summary, it is evident that SN 2019tua initially exhibited distinct hydrogen feature, with later develop strong He I lines (resembling SNe Ib), leading to its classification as a type IIb SN \citep{1997ARA&A..35..309F,2017hsn..book..195G}.
\par To estimate the reddening of the host galaxy, we searched for the interstellar absorption features of the Na ID lines in the corrected spectra. 
Figure \ref{fig:spec2.pdf} reveals no prominent, narrow Na ID absorption lines, implying that ${E(B - V)_{\text{host}}}$ is negligible.
Although utilizing Na ID lines to estimate dust extinction is not entirely reliable—owing to the possibility that the absence of Na ID lines could be attributed to the low resolution of the spectra \citep{2011MNRAS.415L..81P}—the lack of detectable Na ID signatures typically suggests minimal reddening in the host galaxy \citep{2013ApJ...779...38P}. 
Consequently, we infer that the extinction in the host galaxy UGC-11860 is negligible in comparison to the reddening caused by the Milky Way, the total reddening is solely attributed to Galactic sources, applying a correction of $E(B - V)$ = 0.06 mag \citep{2011ApJ...737..103S}.
\begin{figure}[htp]
   \centering
   \includegraphics[width=9cm]{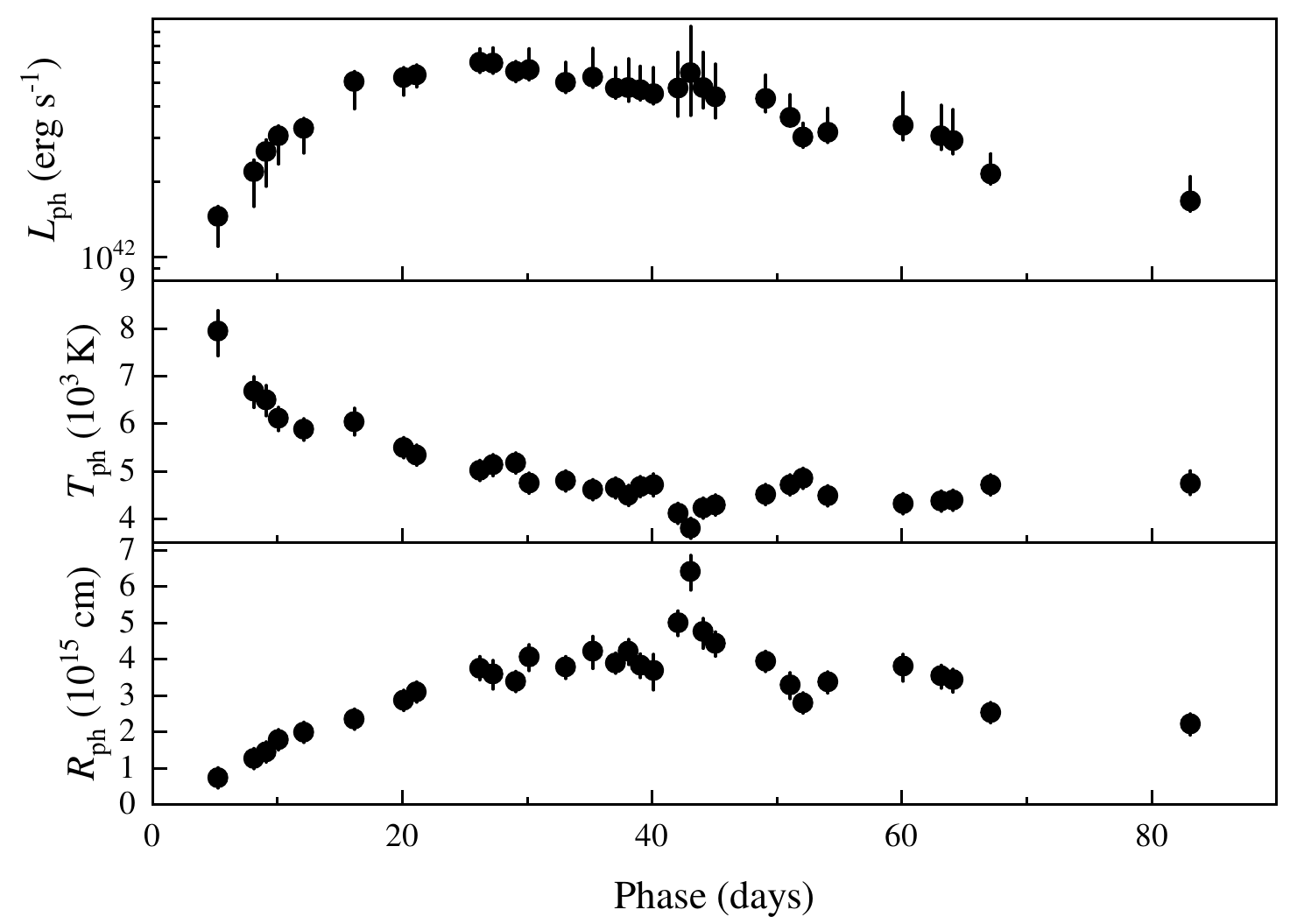}
   \caption{In addition to the main peak, it's very clear to see the bumps on the quasi-bolometric LC. Top panel: The bolometric LC of SN 2019tua, obtained by integrating the flux from the \textit{BVRI} bands. Second panel: effective temperature. Bottom panel: photospheric radius.}
   \label{fig:bol_2019TUA_BVRI.pdf}
\end{figure}
\par Typically, the photospheric velocity of the SN is determined using the absorption linear velocities of Fe II lines at wavelengths 4924 Å, 5018 Å, and 5169 Å \citep{2016A&A...596A..67R}.
During the spectrum of the peaking phase, the velocities for the three Fe II lines are found to be $3700\pm500$ km s$^{-1}$, $4900\pm700$ km s$^{-1}$, and $8100\pm1200$ km s$^{-1}$, respectively. The mean velocity, denoted as $V_{\text{phot}}$, is approximately 5600 km s$^{-1}$ at peak luminosity.
\begin{deluxetable}{ccccc}
\tablecaption{Best-fit Parameters of the Blackbody Fits\label{tab:SED message}}
\tablewidth{0pt}
\tablehead{
\colhead{ Phase } & \colhead{ $L_{\text{ph}}$} & $T_{\text{ph}}$ & $R_{\text{ph}}$ & $ \chi^2 / \text{dof} $ \\
\colhead{ (days) } & ($10^{42}~\mathrm{erg~s^{-1}}$) & ($10^3~\text{K}$) & $(10^{15}~\text{cm})$ & }

\startdata
5.26  &1.45$_{-0.35}^{+0.10}$  &7.94$_{-0.31}^{+0.43}$  & 0.73$_{-0.07}^{+0.07}$&5.50\\
8.13  &2.19$_{-0.60}^{+0.25}$  &6.68$_{-0.34}^{+0.31}$  & 1.26$_{-0.13}^{+0.12}$&4.41\\
9.10  & 2.64$_{-0.72}^{+0.30}$ & 6.47$_{-0.33}^{+0.30}$ & 1.45$_{-0.15}^{+0.14}$&4.48 \\
10.09 & 3.06$_{-0.70}^{+0.25}$ & 6.11$_{-0.26}^{+0.23}$ & 1.78$_{-0.16}^{+0.16}$&5.64 \\
12.12 & 3.27$_{-0.66}^{+0.17}$ & 5.88$_{-0.23}^{+0.22}$ & 1.99$_{-0.18}^{+0.16}$&4.33 \\
16.14 & 5.05$_{-1.13}^{+0.29}$ & 6.05$_{-0.28}^{+0.28}$ & 2.35$_{-0.25}^{+0.22}$&2.30 \\
20.13 & 5.23$_{-0.78}^{+0.02}$ & 5.49$_{-0.18}^{+0.19}$ & 2.87$_{-0.24}^{+0.22}$&2.52\\
21.14 & 5.37$_{-0.56}^{+0.19}$ & 5.54$_{-0.16}^{+0.15}$ & 3.10$_{-0.24}^{+0.22}$&2.35 \\
26.22 & 6.07$_{-0.25}^{+0.79}$ & 5.02$_{-0.17}^{+0.15}$ & 3.75$_{-0.32}^{+0.32}$&4.94 \\
27.26 & 5.97$_{-0.41}^{+0.91}$ & 5.25$_{-0.23}^{+0.21}$ & 3.59$_{-0.41}^{+0.37}$&2.89 \\
29.08 & 5.53$_{-0.33}^{+0.41}$ & 5.18$_{-0.14}^{+0.13}$ & 3.38$_{-0.24}^{+0.23}$&5.50 \\
30.17 & 5.62$_{-0.15}^{+1.21}$ & 4.75$_{-0.15}^{+0.14}$ & 4.06$_{-0.37}^{+0.34}$&6.12 \\
33.09 & 5.01$_{-0.17}^{+1.02}$ & 4.79$_{-0.13}^{+0.14}$ & 3.78$_{-0.31}^{+0.29}$&6.62 \\
35.27 & 5.26$_{-0.33}^{+1.60}$ & 4.61$_{-0.17}^{+0.17}$ & 4.21$_{-0.47}^{+0.41}$&5.14 \\
37.09 & 4.75$_{-0.34}^{+1.00}$ & 4.64$_{-0.11}^{+0.11}$ & 3.89$_{-0.25}^{+0.23}$&4.77 \\
38.09 & 4.76$_{-0.55}^{+1.46}$ & 4.49$_{-0.12}^{+0.13}$ & 4.22$_{-0.37}^{+0.32}$&6.36 \\
39.09 & 4.67$_{-0.27}^{+1.15}$ & 4.69$_{-0.15}^{+0.13}$ & 3.83$_{-0.34}^{+0.31}$&4.80 \\
40.14 & 4.51$_{-0.05}^{+1.24}$ & 4.58$_{-0.23}^{+0.23}$ & 3.69$_{-0.53}^{+0.44}$&2.08 \\
42.08 & 4.74$_{-1.07}^{+1.88}$ & 4.12$_{-0.09}^{+0.09}$ & 5.00$_{-0.35}^{+0.32}$&17.67 \\
43.08 & 5.48$_{-1.79}^{+2.92}$ & 3.80$_{-0.08}^{+0.08}$ & 6.40$_{-0.50}^{+0.45}$&24.22 \\
44.08 & 4.77$_{-0.82}^{+1.84}$ & 4.22$_{-0.12}^{+0.13}$ & 4.75$_{-0.45}^{+0.37}$&4.92\\
45.08 & 4.38$_{-0.77}^{+1.56}$ & 4.29$_{-0.11}^{+0.10}$ & 4.43$_{-0.35}^{+0.31}$&4.66 \\
49.10 & 4.30$_{-0.48}^{+1.06}$ & 4.50$_{-0.09}^{+0.09}$ & 3.94$_{-0.24}^{+0.22}$&12.41 \\
51.08 & 3.62$_{-0.01}^{+0.85}$ & 4.70$_{-0.19}^{+0.20}$ & 3.29$_{-0.38}^{+0.33}$&6.24\\
52.08 & 3.02$_{-0.11}^{+0.41}$ & 4.83$_{-0.17}^{+0.18}$ & 2.79$_{-0.26}^{+0.22}$&6.49 \\
54.09 & 3.16$_{-0.19}^{+0.78}$ & 4.47$_{-0.13}^{+0.14}$ & 3.37$_{-0.03}^{+0.26}$&7.19 \\
60.10 & 3.36$_{-0.42}^{+1.20}$ & 4.31$_{-0.14}^{+0.16}$ & 3.80$_{-0.41}^{+0.33}$&2.59 \\
63.12 & 3.05$_{-0.36}^{+1.00}$ & 4.36$_{-0.12}^{+0.14}$ & 3.54$_{-0.34}^{+0.29}$&6.54 \\
64.11 & 2.92$_{-0.33}^{+0.97}$ & 4.39$_{-0.14}^{+0.13}$ & 3.43$_{-0.34}^{+0.29}$&8.37\\
67.11 & 2.15$_{-0.01}^{+0.44}$ & 4.71$_{-0.16}^{+0.16}$ & 2.53$_{-0.24}^{+0.22}$&8.46 \\
83.11 & 1.67$_{-0.01}^{+0.43}$ & 4.52$_{-0.23}^{+0.27}$ & 2.43$_{-0.30}^{+0.23}$&2.72 \\
\enddata
\end{deluxetable}
\subsection{Bolometric Light Curve} \label{subsec:BOLOMETRIC LIGHT CURVE}
\begin{figure*}
    \centering
    \includegraphics[width=19cm]{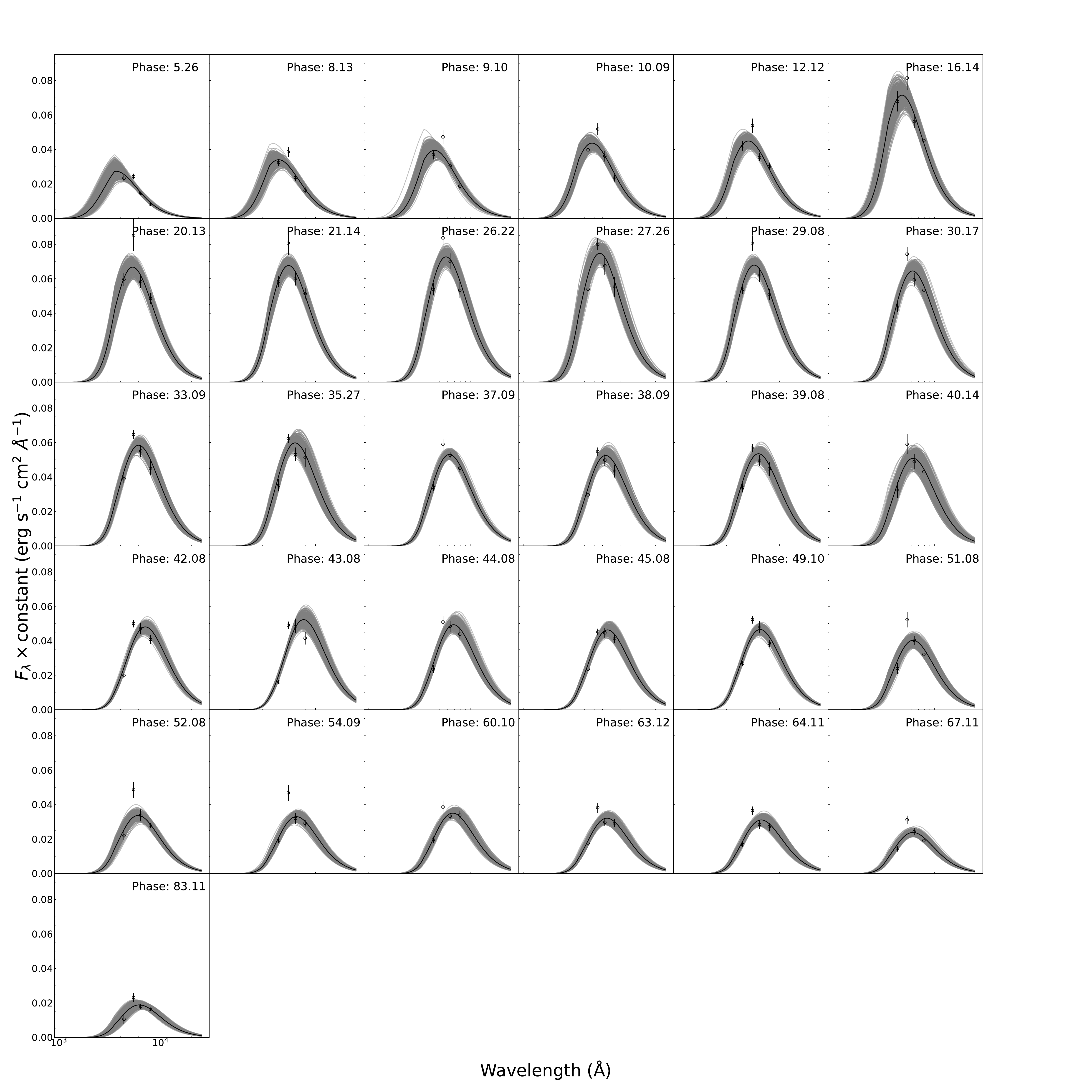}
    \caption{Blackbody fits to the $\textit{BVRI}$ photometry of SN 2019tua. During epochs with no observed in $\textit{BVI}$ bands, the optical data has been interpolated. An excess in the $\textit{V}$ band compared to the blackbody function was observed. The extracted samples are represented by the light gray lines, and the dispersion corresponds to the overall uncertainty in the fit. The black line represents the best fit line associated with the data points.}
    \label{fig:combined_blackbody_fits.pdf}
\end{figure*}

Using \textit{BVRI} photometry, we constructed the spectral energy distribution (SED) and quasi-bolometric LC of SN 2019tua, as well as the effective temperature and photospheric radius.
Due to the limited of optical data, the error in the derived bolometric LC data of SN 2019tua may be slightly larger compared to those derived from SNe with more bands.
The fitting procedure for the bolometric LC employs the publicly accessible \textit{Superbol} program \citep{2018zndo...2155821M}.

After conducting reddening corrections on the collected data and adjusting for the distance modulus at \( z = 0.010 \), linear interpolation is executed on each band of the LC to supplement missing data points with corresponding observations in the \textit{R} band. 
And then, spectral energy distributions (SEDs) are generated for each point on the LC using the \textit{emcee} Python package \citep{2013PASP..125..306F}. 
Each spectral energy distribution is then fitted with a blackbody function via the nonlinear least squares method, using the \textit{curve-fit} function in \textit{scipy} \citep{2020NatMe..17..261V}. The uncertainties are estimated using a Markov Chain Monte Carlo (MCMC) approach. 
We ran the MCMC with 600 steps
The results of the fitting are illustrated in Figure \ref{fig:bol_2019TUA_BVRI.pdf}, and the best-fit parameters are enumerated in Table \ref{tab:SED message} and depicted in Figure \ref{fig:combined_blackbody_fits.pdf}. 
We employed the blackbody equation to fit the SEDs of SN 2019tua, thereby deriving the bolometric LC, as well as the temperature and radius evolution for each epoch of SN 2019tua. 

\par Taking into account the observed excess in the \textit{V} band within SN 2019tua's SED, particularly in conjunction with the time corresponding to the bump in the LCs in \ref{fig:lcs}, a sustained excess is evident in the \textit{V} band from approximately day 40 to day 65.
It could be the presence of a transient additional energy source in the \textit{V} band. Such an ancillary radiation mechanisms could significantly contribute to the observed excess in this band \citep{2017ApJ...840...57Y,2017ApJ...835L...8N}. 
From Figure \ref{fig:lcs}, the performance of the \textit{B} band appears inconsistent when compared to other bands. 
Less pronounced fluctuations are observed in other bands than in the \textit{B} band. These variations among the bands have implications for the quality of our SED fit, and subsequently influence the derived bolometric LC.
We conducted observations of SN 2019tua across multiple bands from day 40 to day 55, during which a noticeable bump was evident in the quasi-bolometric LC. 
The secondary peak in the \textit{B} band is more pronounced compared to other bands, as depicted in Figure \ref{fig:lcs}. 
This situation may be the reason for the maximum error in SED fitting to occur at around 40 days, thereby amplifying the error of the bolometric LC at the same time.

\begin{figure}[htp]
   \centering
   \hspace{-1.4cm}
   \includegraphics[width=9.5cm]{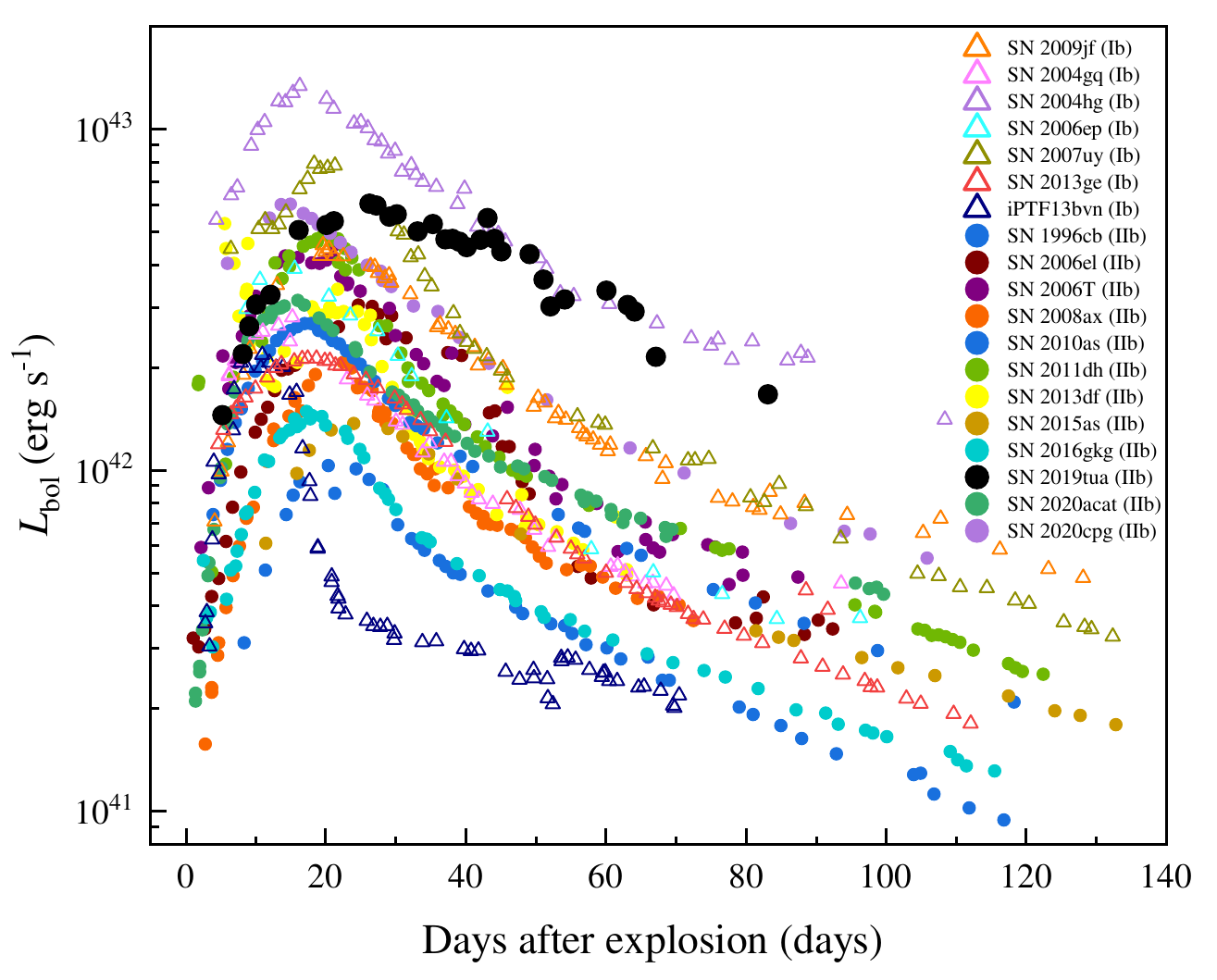}
   \caption{Bolometric light curves of SN 2019tua in comparison with some of other type IIb and type Ib SNe. Type IIb SNe are shown as circles, while 
 hollow triangular markers represent type Ib SNe. }
   \label{fig:refbol.pdf}
\end{figure}
\begin{figure}[htp]
   \centering
   \hspace{-1.4cm}
   \includegraphics[width=9.5cm]{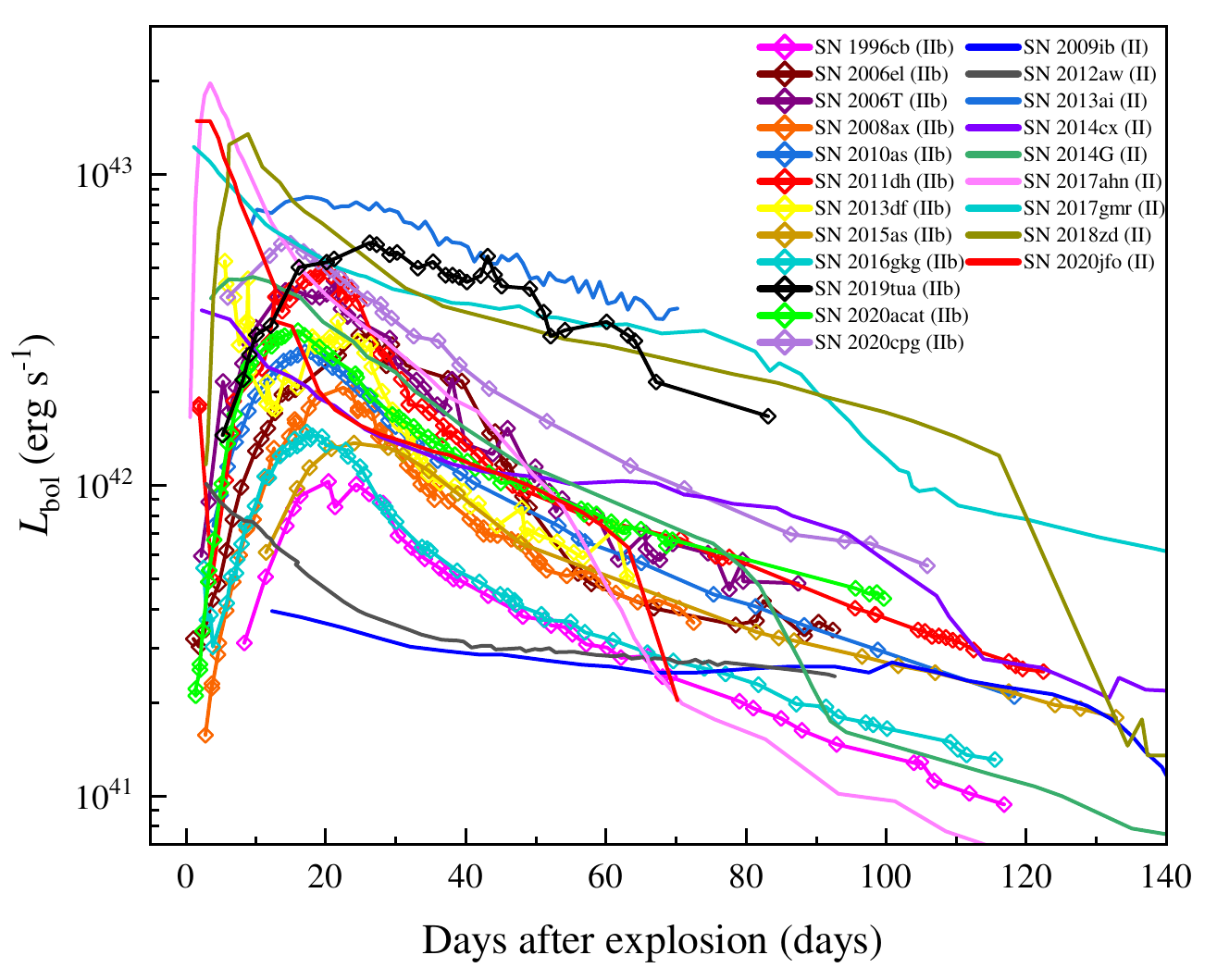}
   \caption{Bolometric light curves of SN 2019tua and some type IIb SNe in comparison with some of other type II SNe. Type IIb SNe are shown by colored hollow rhombus connected by colored solid lines and the other type II SNe are shown by colored solid lines.}
   \label{fig:refbol1.pdf}
\end{figure}

\par 
We determined the peak time ($t_{\text{peak}}$) of SN 2019tua by directly selecting the period with the maximum luminosity value from Table \ref{tab:SED message}, yielding $t_{\text{peak}}$ = $26.22 \pm 1.0$ days. Consequently, we obtained the peak luminosity $L_{\text{peak}}$ of SN 2019tua is \(6.07~\times 10^{42} \, \text{erg s}^{-1}\), substantially exceeding the typical luminosity of SNe IIb and even surpassing the $L_{\text{peak}}$ of some SNe Ib \citep{2021ApJ...918...89A} (in Figure \ref{fig:refbol.pdf}). 
And the bolometric LC of SN 2019tua is consistent with the general trend illustrated in Figure \ref{fig:lcs}.
Those bolometric LCs were obtained from sne.space and used the same method as described in the \citep{2021ApJ...918...89A}, and the explosion times used were also derived from \citep{2021ApJ...918...89A}.

\begin{figure*}
    \centering
    \hspace{0cm}
    \includegraphics[width=16cm]{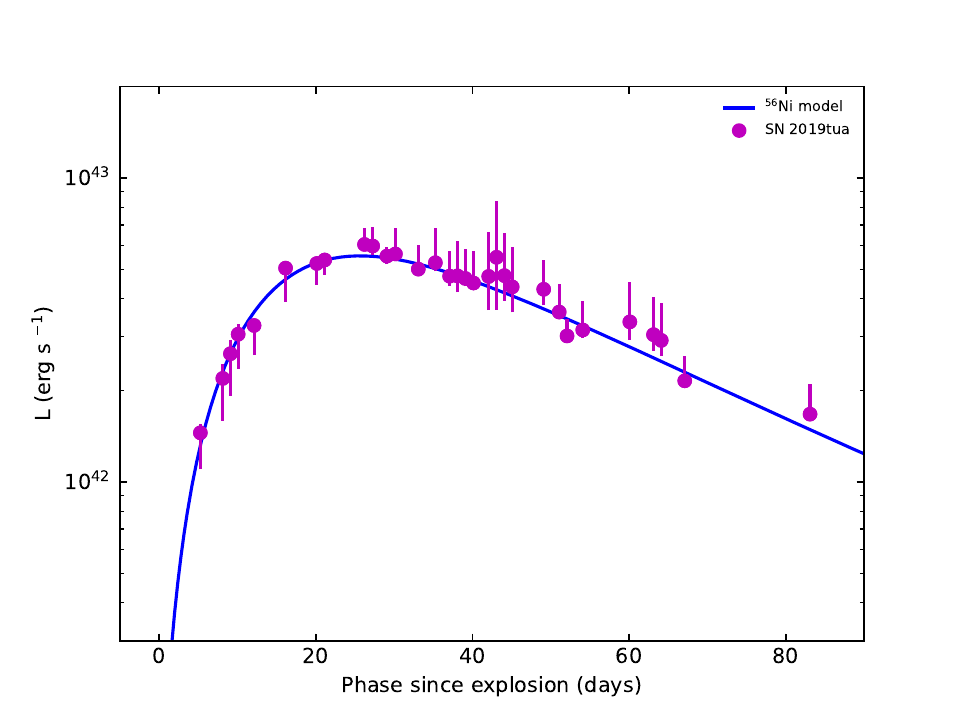}
    \caption{The bolometric LC of SN 2019tua reproduced by the $^{56}$Ni model. The total luminosity $L_{\mathrm{\rm inp,Ni}}$ is represented by the solid blue line.}
    \label{fig:56Ni.pdf}
\end{figure*}
\begin{deluxetable*}{cccccccc}
\tablecaption{Parament of the $^{56}$Ni Model\label{tab:Ni}}
\tablewidth{0pt}
\tablehead{
         \colhead{ $M_{\text{ej}}$ } & \colhead{$M_{\text{Ni}}$ }&\colhead{$\kappa$} & \colhead{$\kappa_{\gamma, \mathrm{Ni}}$}&\colhead{$t_{\text{expl}}$}& \colhead{ $ \chi^2 / \text{dof} $ } &\colhead{AIC}& \colhead{BIC}\\
 \colhead{($M_{\odot}$)} & \colhead{($M_{\odot}$)}&\colhead{(cm$^{2}$~g$^{-1}$)} & \colhead{(cm$^{2}$~g$^{-1}$)} & \colhead{(days)}& \colhead{ } & \colhead{ }& \colhead{ }
}
\startdata
1.38$_{-0.57}^{+0.99}$ & 0.38$_{-0.02}^{+0.03}$ & 0.19$_{-0.08}^{+0.13}$ & 0.01$_{-0.00}^{+0.01}$ & -1.05$_{-0.87}^{+0.80}$ & 0.37 & 19.65 & 55.70 \\
\enddata
\tablecomments{The uncertainties are $1\sigma$.}
\end{deluxetable*}
\begin{figure*}
    \centering
    \includegraphics[angle=0,scale=0.60]{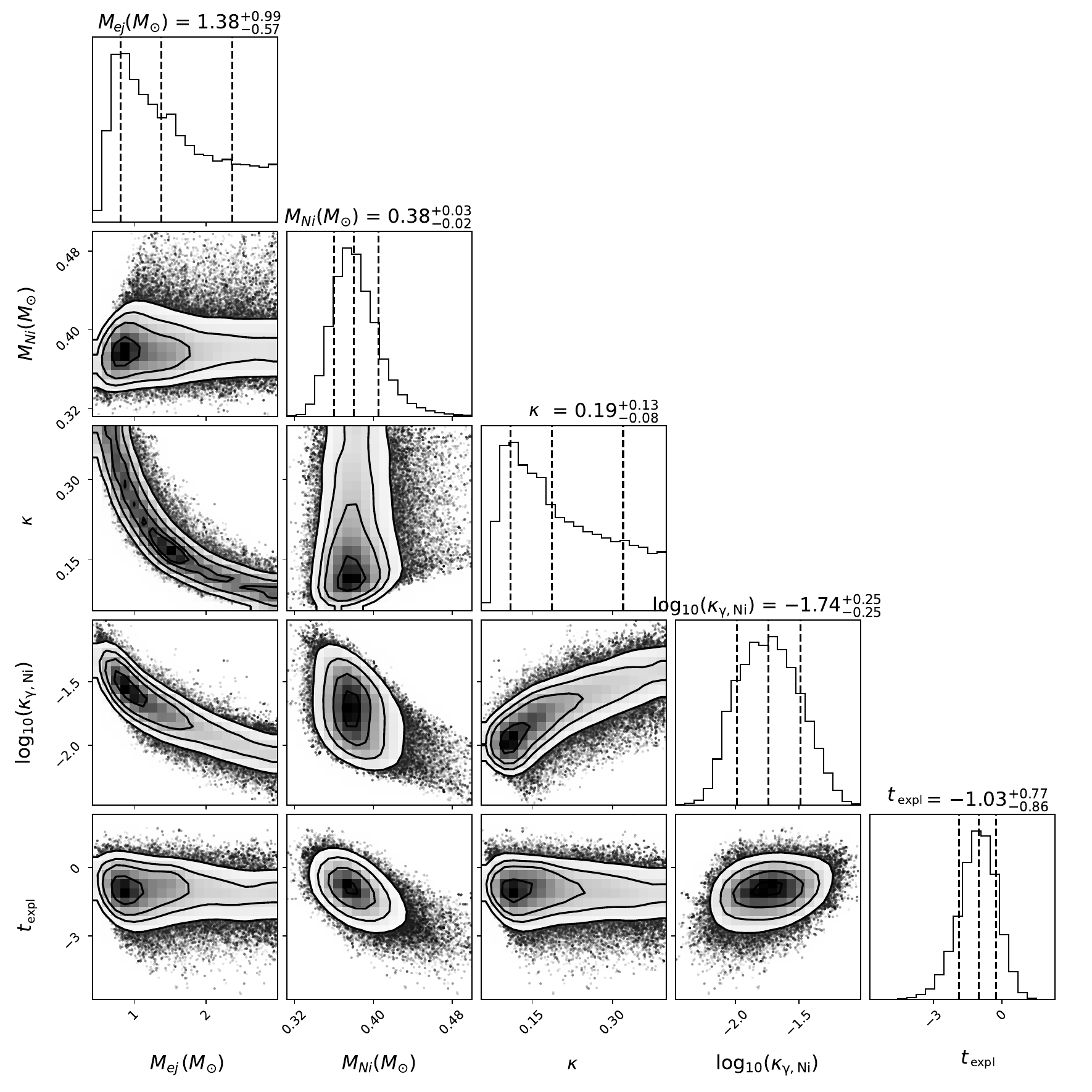}
    \caption{The corner plot for the $^{56}$Ni model. The uncertainties are computed as the 16th and 84th percentiles (1$\sigma$) of the posterior samples along each axis.}
    \label{fig:56Ni_corner.pdf}
\end{figure*}

Additionally, SN 2019tua exhibits higher luminosity compared to some other type IIb SNe, such as SN 2010as $\sim$ \(2.7~\times 10^{42} \, \text{erg s}^{-1}\) \citep{2014ApJ...792....7F}, SN 2020acat $\sim$ \(3.1~\times 10^{42} \, \text{erg s}^{-1}\) \citep{2022MNRAS.513.5540M,2024A&A...683A.241E}, SN 2015as $\sim$ \(5.2~\times 10^{42} \, \text{erg s}^{-1}\) \citep{2018MNRAS.476.3611G}.
And the $L_{\text{peak}}$ of SN 2019tua is similar to SN 2020cpg $\sim$ \(6.03~\times 10^{42} \, \text{erg s}^{-1}\) \citep{2021MNRAS.506.1832M,2022MNRAS.517.5678T}.
We also plotted the LCs from the reference \citep{2014ApJ...792....7F,2022MNRAS.513.5540M,2022MNRAS.517.5678T,2018MNRAS.476.3611G} as shown in Figure \ref{fig:refbol.pdf}, with the aligned the explosion time.
In contrast, the luminosity of SN 2019tua is lower than the type IIb, such as SN 2018gk  $\sim$ \(7.0~\times 10^{43} \, \text{erg s}^{-1}\) \citep{2021MNRAS.503.3472B}.

\par In addition, we further attempted to compare the bolometric LCs of SN 2019tua and some type IIb SNe with those of other well-documented type II SNe, such as SN 2009ib \citep{2015MNRAS.450.3137T}, SN 2012aw \citep{2014ApJ...782...30Y}, SN 2013ai \citep{2021ApJ...909..145D}, SN 2014G \citep{2016MNRAS.462..137T}, SN 2014cx \citep{2016ApJ...832..139H}, SN 2017ahn \citep{2021ApJ...907...52T}, SN2017gmr \citep{2019ApJ...885...43A}, SN 2018zd \citep{2020MNRAS.498...84Z}, SN 2022jfo \citep{2022ApJ...930...34T}. 
The results are presented in Figure \ref{fig:refbol1.pdf} and we also plotted these LCs from the reference with the aligned the explosion time, respectively. 
It can be clearly observed that, compared to type II SNe, type IIb SNe generally take a longer time to rise to their peak luminosities.
The type II SNe typically reach peak luminosity within one to two weeks, whereas type IIb SNe require more time \citep{2019MNRAS.488.4239P}.  
Additionally, the late-time luminosity decline of type IIb SNe is noticeably slower compared to type II-L SNe but faster than type II-P SNe and the different decline rates were proposed to correlate with the amount of hydrogen retained by the progenitor \citep{2021ApJ...909..145D}.
It has been claimed that type Ib, IIb, II-L, and type II-P SNe
all result from the collapse of massive stars, with
their main differences depending on the details of the mass loss and the amount of mixing \citep{2021ApJ...909..145D}.

\begin{figure*}
    \centering
     \hspace{0cm}
    \includegraphics[width=15cm]{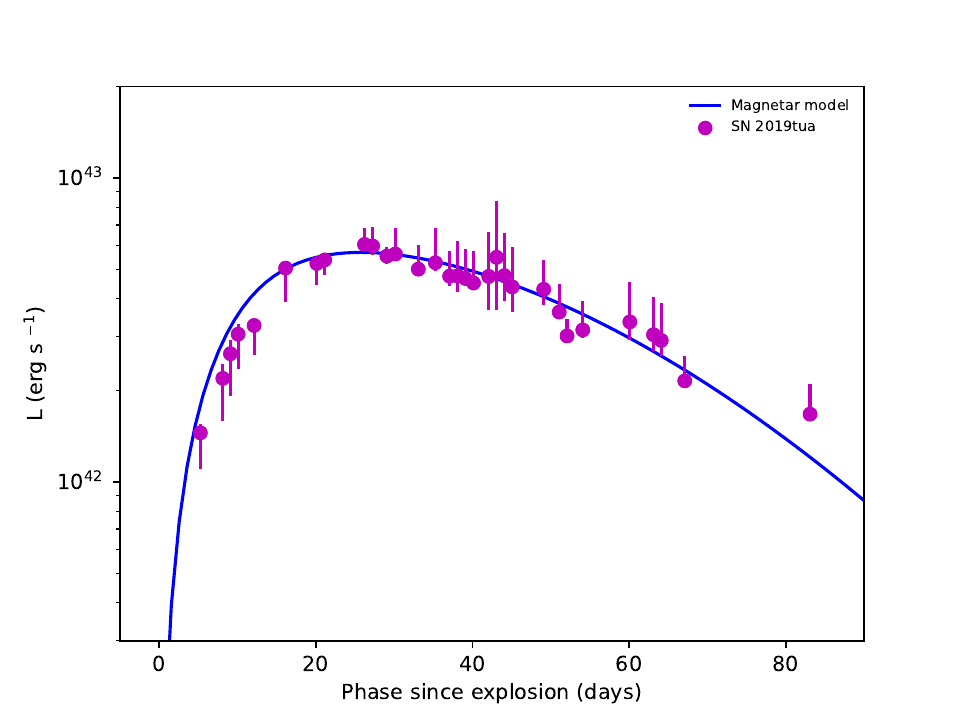}
    \caption{The bolometric LC of SN 2019tua reproduced by the magnetar model. The total luminosity $L_{\mathrm{inp,mag}}$ is represented by the solid blue line.}
    \label{fig:mag.pdf}
\end{figure*}
\begin{deluxetable*}{ccccccccc}
\tablecaption{Parament of the Magnetar Model\label{tab:Magnetar}}
\tablewidth{0pt}
\tablehead{\colhead{ $M_{\text{ej}}$ } & \colhead{$P_{\text{NS}}$}&\colhead{$\kappa$} &\colhead{$B$} & \colhead{$\kappa_{\gamma, \mathrm{mag}}$} &\colhead{ $t_{\text{expl}}$ } & \colhead{ $ \chi^2 / \text{dof} $ }&\colhead{AIC}& \colhead{BIC} \\
 \colhead{($M_{\odot}$)} & \colhead{(ms)}&\colhead{(cm$^{2}$~g$^{-1}$)} & \colhead{(10$^{14}$~G)} & \colhead{(cm$^{2}$~g$^{-1}$)} & \colhead{(days)}& \colhead{ }& \colhead{} & \colhead{}}
\startdata
3.81$_{-1.68}^{+2.51}$ & 9.66$_{-0.87}^{+0.98}$ & 0.39$_{-0.16}^{+0.30}$ & 9.50$_{-2.58}^{+2.81}$ & 0.01$_{-0.01}^{+0.00}$ & -0.47$_{-0.94}^{+1.28}$ & 0.98 & 36.45 & 79.71\\
\enddata
\tablecomments{The uncertainties are $1\sigma$.}
\end{deluxetable*}
\begin{figure*}
    \centering
    \includegraphics[angle=0,scale=0.5]{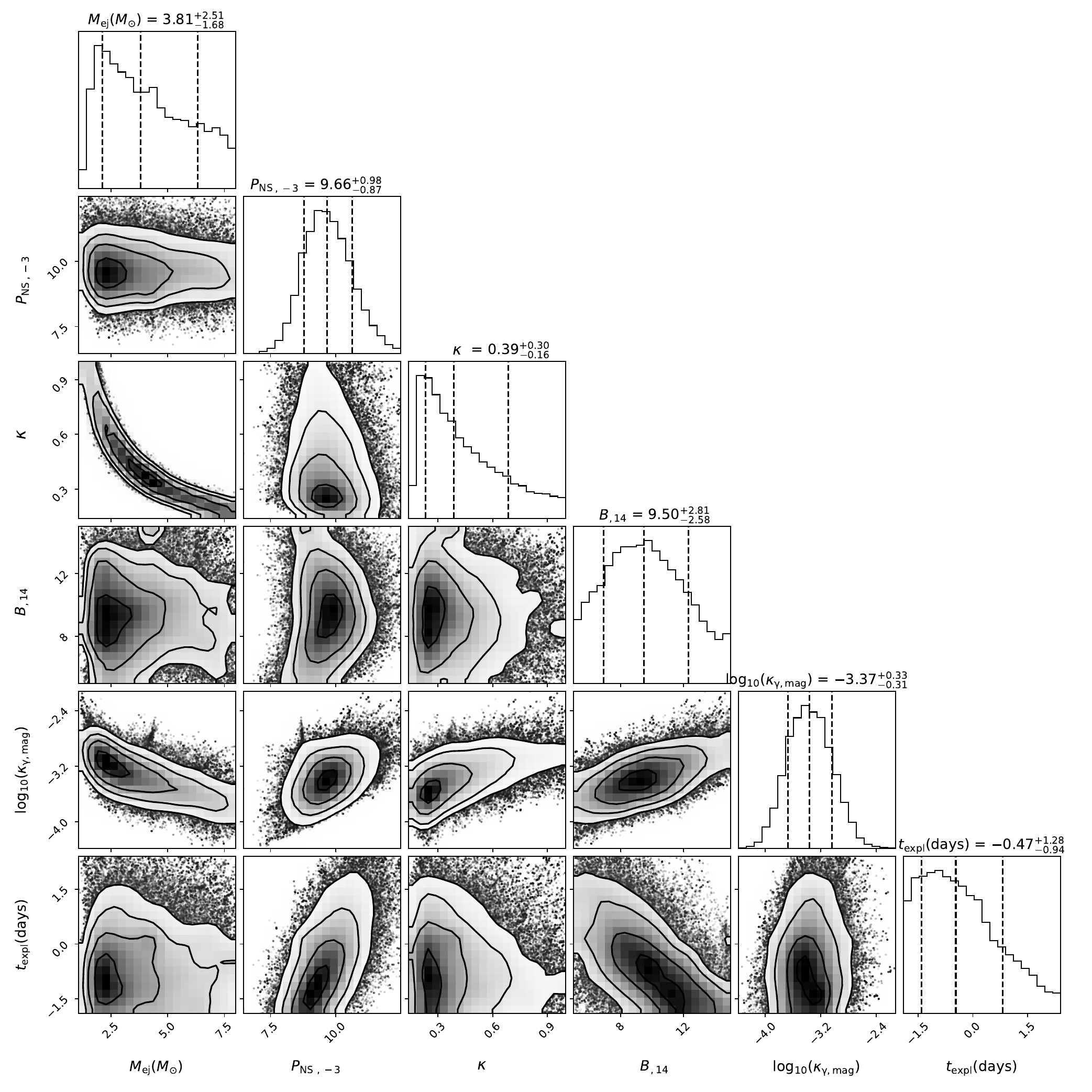}
    \caption{The corner plot for the magnetar model. The uncertainties are computed as the 16th and 84th percentiles (1$\sigma$) of the posterior samples along each axis.}
    \label{fig:mag_corner.pdf}
\end{figure*}

\section{Modelling of SN 2019tua} \label{sec:Modelling of SN 2019tua}
In this section, we apply four distinct energy source models to fit the bolometric LC of SN 2019tua: the \(^{56}\text{Ni}\) model, the ejecta-circumstellar medium interaction (CSI) model, the magnetar model, and the combined CSI plus \(^{56}\text{Ni}\) model.
We restrict our focus to dense shell CSM models with \(s = 0\).
To identify the optimal fit parameters, we employ Bayesian analysis using the \textit{emcee} Python package \citep{2013PASP..125..306F}, which is based on MCMC techniques. Maximum likelihood \( \chi^2 \) fitting is performed, and posterior probability distributions for the free parameters are provided.
To assess the quality of the fits, we conduct Akaike and Bayesian Information Criterion (AIC/BIC) tests.
Details of free parameters and their prior conditions for each model are elaborated in the respective tables.
Simulations are executed with 100,000 steps, and the fitting uncertainty is quantified within a 1\(\sigma\) confidence interval.

In the case of a homogeneously expanding photosphere with energy injection concentrated at the center, the observed bolometric luminosity can be expressed as follows \citep{1982ApJ...253..785A,2012ApJ...746..121C}:

\begin{equation}
L_{\rm bolo}(t)=\frac{2}{\tau_{m}}e^{-\frac{t^{2}}{\tau_{m}^{2}}} \int_0^t e^{\frac{t'^{2}}{\tau_{m}^{2}}} \frac{t'}{\tau_{m}}L_{\rm inp}(t')dt',
\end{equation}
where $L_{\rm inp}(t')$ is the input luminosity, $\tau_m=\left(\frac{10 \kappa M_{\mathrm{ej}}}{3 \beta v c}\right)^{1 / 2}$ is the effective LC timescale, $\kappa$ is the gray opacity, and $\beta \simeq 13.4$
is a dimensionless constant related to the geometry of the ejecta. The ejecta velocity is set to \(v = 5600\) km s\(^{-1}\) based on the velocities of the Fe II lines derived in the previous section.

\begin{figure*}
    \centering
    \includegraphics[width=16cm]{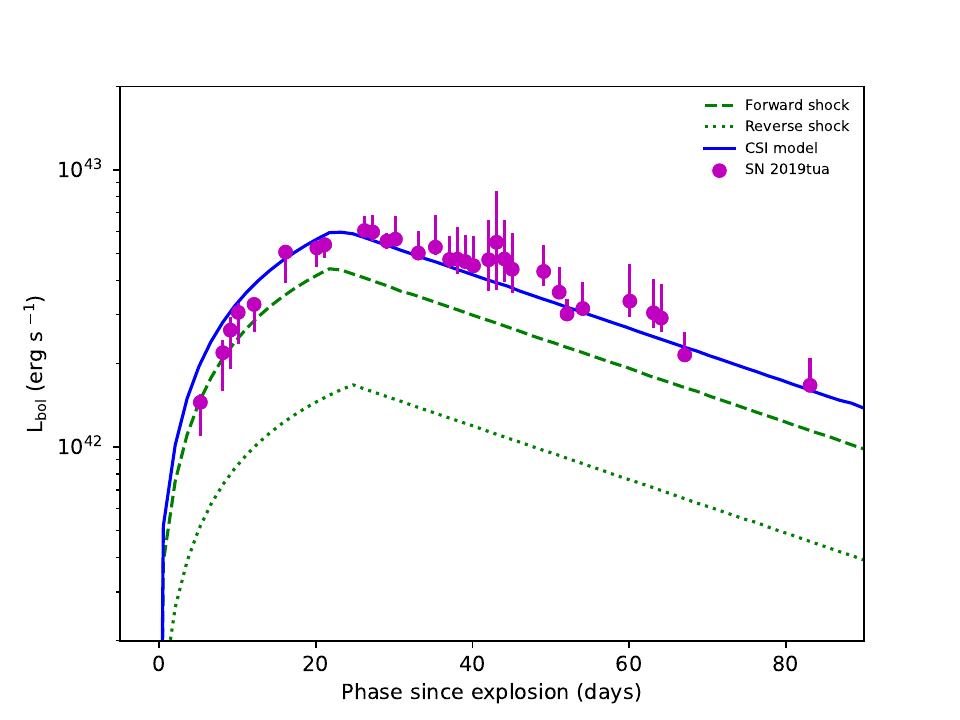}
    \caption{The bolometric LC of SN 2019tua reproduced by the CSI model. The total luminosity $L_{\mathrm{CSI}}$ is represented by the solid blue line, the contribution of the forward shock $L_{\mathrm{inp}, \mathrm{FS}}$ is shown by the green dashed line, and the contribution of the reverse shock $L_{\mathrm{inp}, \mathrm{RS}}$ is shown by the green dotted line.}
    \label{fig:CSI-single.pdf}
\end{figure*}
\begin{deluxetable*}{ccccccccccccc}
\tablecaption{Parament of the CSI Model\label{tab:CSM}}
\tablewidth{0pt}
\tablehead{
         \colhead{ $M_{\text{ej}}$ } & \colhead{$M_{\text{CSM}}$}&\colhead{$\kappa$} &\colhead{$\rho_{\text{CSM}}$} & \colhead{$R_{\text{CSM,in}}$} &\colhead{$\epsilon $} & \colhead{ $x_{\text{0}}$ }&\colhead{ $t_{\text{expl}}$ } & \colhead{ $ \chi^2 / \text{dof} $ } &\colhead{AIC}& \colhead{BIC}\\
 \colhead{($M_{\odot}$)} & \colhead{($M_{\odot}$)}&\colhead{(cm$^{2}$ g$^{-1}$)}&\colhead{( 10$^{-12}$ g cm$^{-3}$)} & \colhead{(10$^{14}$ cm)} & \colhead{} & \colhead{}& \colhead{ (days)} &\colhead{}& \colhead{} & \colhead{}
}
\startdata
2.44$_{-1.20}^{+1.76}$ & 4.19$_{-1.07}^{+1.04}$ & 0.25$_{-0.11}^{+0.10}$ & 28.83$_{-11.38}^{+10.91}$ & 14.02$_{-6.10}^{+6.00}$ & 0.50$_{-0.16}^{+0.20}$ & 0.34$_{-0.05}^{+0.07}$ & -0.98$_{-0.87}^{+0.94}$ & 0.35 & 23.98& 81.66\\
\enddata
\tablecomments{The uncertainties are $1\sigma$.}
\end{deluxetable*}
\begin{figure*}
    \centering
    \includegraphics[angle=0,scale=0.4]{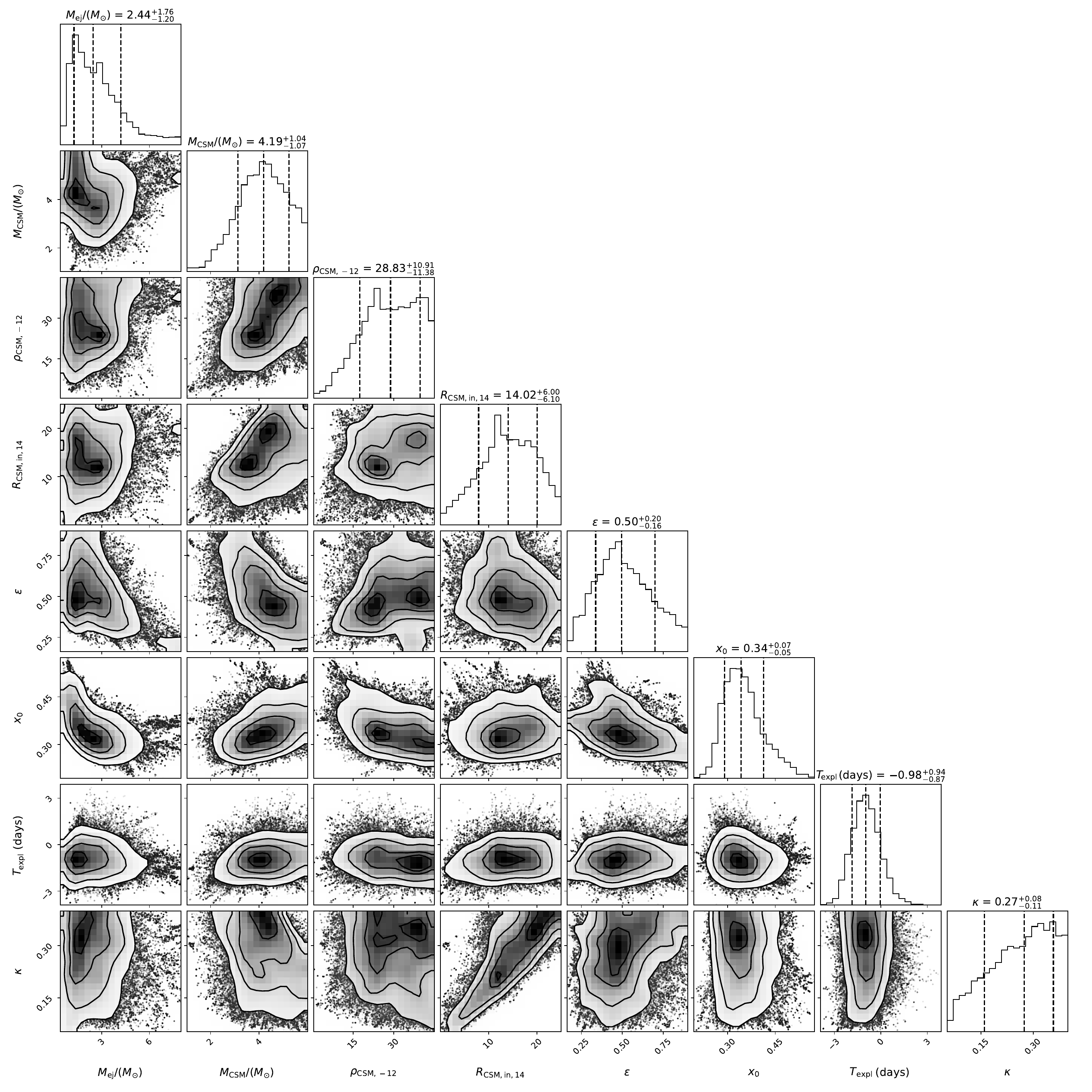}
    \caption{The corner plot for the CSI model ($s = 0$). The uncertainties are computed as the 16th and 84th percentiles (1$\sigma$) of the posterior samples along each axis.}
    \label{fig:CSI-corner.pdf}
\end{figure*}

\subsection{The $^{56}$Ni Model } \label{the $^{56}$Ni modelling}
The input luminosity produced by the decay chain $^{56}$Ni$\rightarrow^{56}$Co$\rightarrow^{56}$Fe can be written as \citep{1969ApJ...157..623C,1980ApJ...237L..81C,1980ApJ...237..541A,1982ApJ...253..785A}:
\begin{equation}
\begin{aligned}
&L_{\rm inp,Ni}(t) =M_{\rm Ni}  \\
&\times \left[\left( \epsilon _{%
\mathrm{Ni}}-\epsilon _{\mathrm{Co}}\right) e^{-t/\tau _{\mathrm{Ni}%
}}+\epsilon _{\mathrm{Co}}e^{-t/\tau _{\mathrm{Co}}}\right]\left(1-\mathrm{e}^{-\tau_{\gamma,\rm{Ni}}}\right),
\end{aligned}
\end{equation}
where $M_{\rm Ni}$ is the mass of $^{56}$Ni. $\epsilon _{\mathrm{Ni}%
}=3.9\times 10^{10}~\rm{erg}~\rm{g}^{-1}~\rm{s}^{-1}$ and $\epsilon _{%
\mathrm{Co}}=6.78\times 10^{9}~\rm{erg}~\rm{g}^{-1}~\rm{s}^{-1}$. $\tau _{%
\mathrm{Ni}}=8.8~\rm days$ and $\tau _{\mathrm{Co}}=111.3~\rm days$ are the decay time of $^{56}$Ni decays to $^{56}$Co and $^{56}$Co decays to $^{56}$Fe, respectively. 
$\tau_{\gamma,\rm{Ni}}$ is the optical depth of the ejecta to gamma-rays and can be written as $\tau_{\gamma,\rm{Ni}}=3 \kappa_{\gamma,\rm{Ni}} M_{\mathrm{ej}} / 4 \pi R_{\mathrm{ej}}^2$.

\par The \(^{56}\text{Ni}\) model has five free parameters: the optical opacity \(\kappa\) (cm$^{2}$~g$^{-1}$), the ejecta mass \(M_{\text{ej}}\) ($M_{\odot}$), the \(^{56}\text{Ni}\) mass \(M_{\text{Ni}}\) ($M_{\odot}$), the gamma-ray opacity of \(^{56}\text{Ni}\) decay photons \(\kappa_{\gamma}\) (cm$^{2}$~g$^{-1}$), and the epoch of the explosion $t_{\text{expl}}$ (days). The fitting results are illustrated in Figure \ref{fig:56Ni.pdf} and summarized in Table \ref{tab:Ni}. 
The corner plots of the \(^{56}\text{Ni}\) model are shown in Figure \ref{fig:56Ni_corner.pdf}.

\subsection{The Magnetar Model} \label{the Magnetar modelling}

\par The magnetar model accounts for the LC of SNe by converting the rotational energy of a newborn magnetar—originating from the collapse of a massive star—into the thermal energy of the SN ejecta. The input power generated through magnetic dipole radiation can be described by
\begin{equation}
L_{\rm inp,{\rm mag}}(t) = \frac{E_{p}}{\tau_{p}} \frac{1}{(1+t/\tau_{p})^{2}}\left(1-e^{-At^{-2}}\right),
\label{equ:input-mag}
\end{equation}
where $E_{p}\simeq~({1}/{2})I_{\rm NS}\Omega^2$ is the rotational energy of a magnetar,
$\tau_{p}~=~{6I_{\rm NS} c^3}/{B^2R_{\rm NS}^6\Omega^2}$ is the spin-down timescale of the magnetar, $I_{\rm NS}~=(2/5)M_{\rm NS}R_{\rm NS}^{2}$ is the moment of inertia of a magnetar \citep{2010ApJ...717..245K}. $M_{\rm NS}=1.4~M_\odot$, $R_{\rm NS}=10$ km and $P=2\pi/\Omega$ are the mass, radius and rotational period of the magnetar, respectively.
A trapping factor $1-e^{-At^{-2}}$ is introduced in the magnetar-powered model to represent gamma/X-ray leakage and trapping rate \citep{2015ApJ...807..147W}, where $A$ can be written as $A=\frac{3\kappa_{\gamma,\rm{mag}} M_{\rm ej}}{4\pi v^2 }$.
\par In the magnetar model, we consider six free parameters for fitting: the ejected mass $M_{\mathrm{ej}}$ ($M_{\odot}$), the initial spin period of the magnetar $P_{\mathrm{NS}}$ (ms), the optical opacity $\kappa$ (cm$^{2}$ g$^{-1}$), the magnetic field strength $B$ (10$^{14}$ G), the gamma-ray opacity for photons generated by magnetar spin-down $\kappa_{\mathrm{\gamma,mag}}$ (cm$^{2}$ g$^{-1}$), and the explosion time $t_{\mathrm{expl}}$ (days). The theoretical bolometric LC fitted using the magnetar model is depicted in Figure \ref{fig:mag.pdf}, and the best-fit parameters are listed in Table \ref{tab:Magnetar}.
The corner plots of the magnetar model are shown in Figure \ref{fig:mag_corner.pdf}.

\begin{figure*}
    \centering
    \includegraphics[width=16cm]{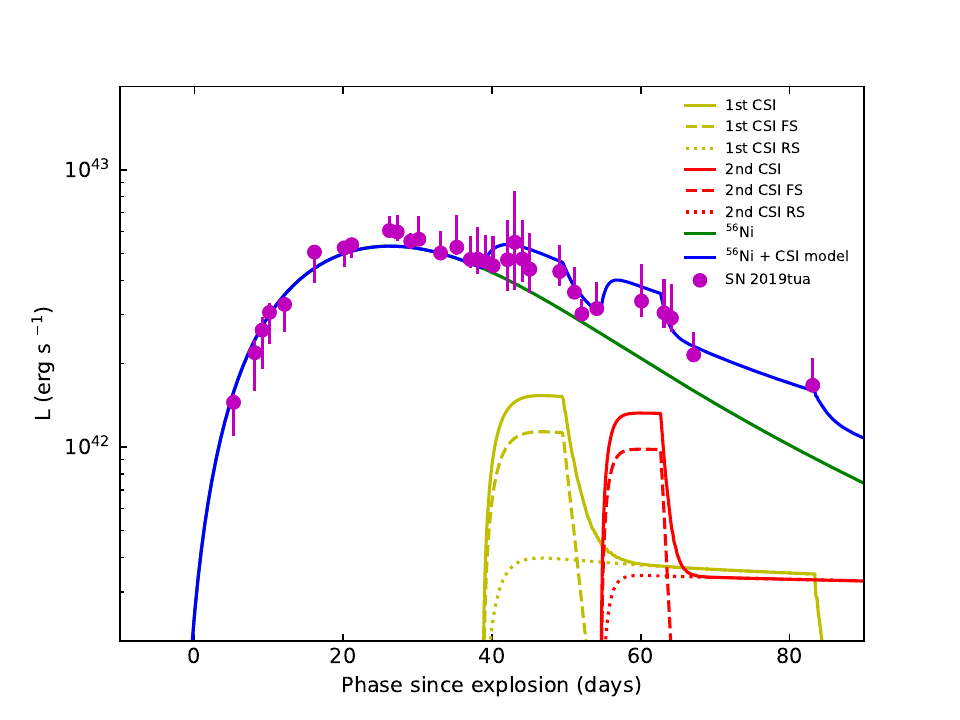}
    \caption{The bolometric LC of SN 2019tua reproduced by the double CSI plus the $^{56}$Ni model. The total luminosity $L_{bolo}$ is represented by the solid blue curve, and the contribution of $^{56}$Ni is represented by the solid yellow line. The contributions of the two consecutive CSI events are shown as different colors of dashed and dotted curves, with green representing the contribution of the first CSI event and red representing the contribution of the second CSI event. The contribution of the forward shock $L_{\mathrm{inp,FS,\mathit{i}}}$ in the CSI event is represented by a dashed line, while the contribution of the reverse shock $L_{\mathrm{inp,RS,\mathit{i}}}$ is shown as a dotted line.}
    \label{fig:CSI+56Ni.pdf}
\end{figure*}

\setlength{\tabcolsep}{1.5mm}{
\begin{table*}[htbp]
\footnotesize
\centering
\tabcolsep=0.0cm
\tablewidth{0pt}
\caption{Parament of the CSI Plus $^{56}$Ni Model}
\begin{tabular}{cccccccccccccccc}
\hline
\hline
\colhead{ $M_{\mathrm{Ni}}$ } & \colhead{${\kappa_{\gamma,\mathrm{Ni}}}$}&\colhead{Interaction} &\colhead{ $s$} & \colhead{$\kappa$} & \colhead{$M_{\mathrm{ej,}\textit{i}}$ } & \colhead{ $M_{\mathrm{CSM},\textit{i}}$ } & \colhead{$R_{\mathrm{CSM,in,}\textit{i}}$ } & \colhead{ $\epsilon$$_i$ } & \colhead{ $\rho_{\mathrm{CSM,in,}\textit{i}}$ } &\colhead{ $x_{\mathrm{0}}$ } &\colhead{ $t_{\mathrm{expl}}$ } &\colhead{ $t_{\mathrm{CSI,}\textit{i}}$ } & \colhead{ $ \chi^2 / \mathrm{dof} $ } &\colhead{AIC}& \colhead{BIC}\\
\colhead{($M_{\odot}$)} & \colhead{($\rm cm^{2}~g^{-1}$)}&\colhead{$i$th} & \colhead{} & \colhead{($\rm cm^{2}~g^{-1}$)} & \colhead{($M_{\odot}$)} & \colhead{($M_{\odot}$)} & \colhead{($10^{14}$~cm)} & \colhead{} &\colhead{(10$^{-12}$ g cm$^{-3}$)} & \colhead{}  & \colhead{(days)}& \colhead{ (days)}& \colhead{} & \colhead{}\\
\hline
0.41$_{-0.01}^{+0.01}$  & 0.01$_{-0.01}^{+0.00}$ & 1 & 0 & 0.17$_{-0.00}^{+0.01}$ & 2.03$_{-0.02}^{+0.05}$ & 0.38$_{-0.02}^{+0.06}$& 7.10$_{-2.09}^{+2.18}$ & 0.32$_{-0.01}^{+0.03}$ &29.26$_{-2.58}^{+3.27}$ & 0.18$_{-0.00}^{+0.01}$& -2.65$_{-0.67}^{+0.45}$& 38.62$_{-1.44}^{+1.15}$ & 0.85 & 43.69& 151.85 \\
&& 2 & 0 & & 2.41 & 0.63$_{-0.09}^{+0.15}$ & $14.81$ & 0.24$_{-0.03}^{+0.06}$ & 49.30$_{-9.49}^{+4.99}$  &  &   & 54.55$_{-1.82}^{+2.29}$ & &  &  \\
\hline
\end{tabular}
\label{tab:duoble+ni}
\begin{flushleft}
\tablecomments{
The uncertainties are $1\sigma$.
}
\end{flushleft}
\end{table*}}

\begin{figure*}
    \centering
    \includegraphics[angle=0,scale=0.22]{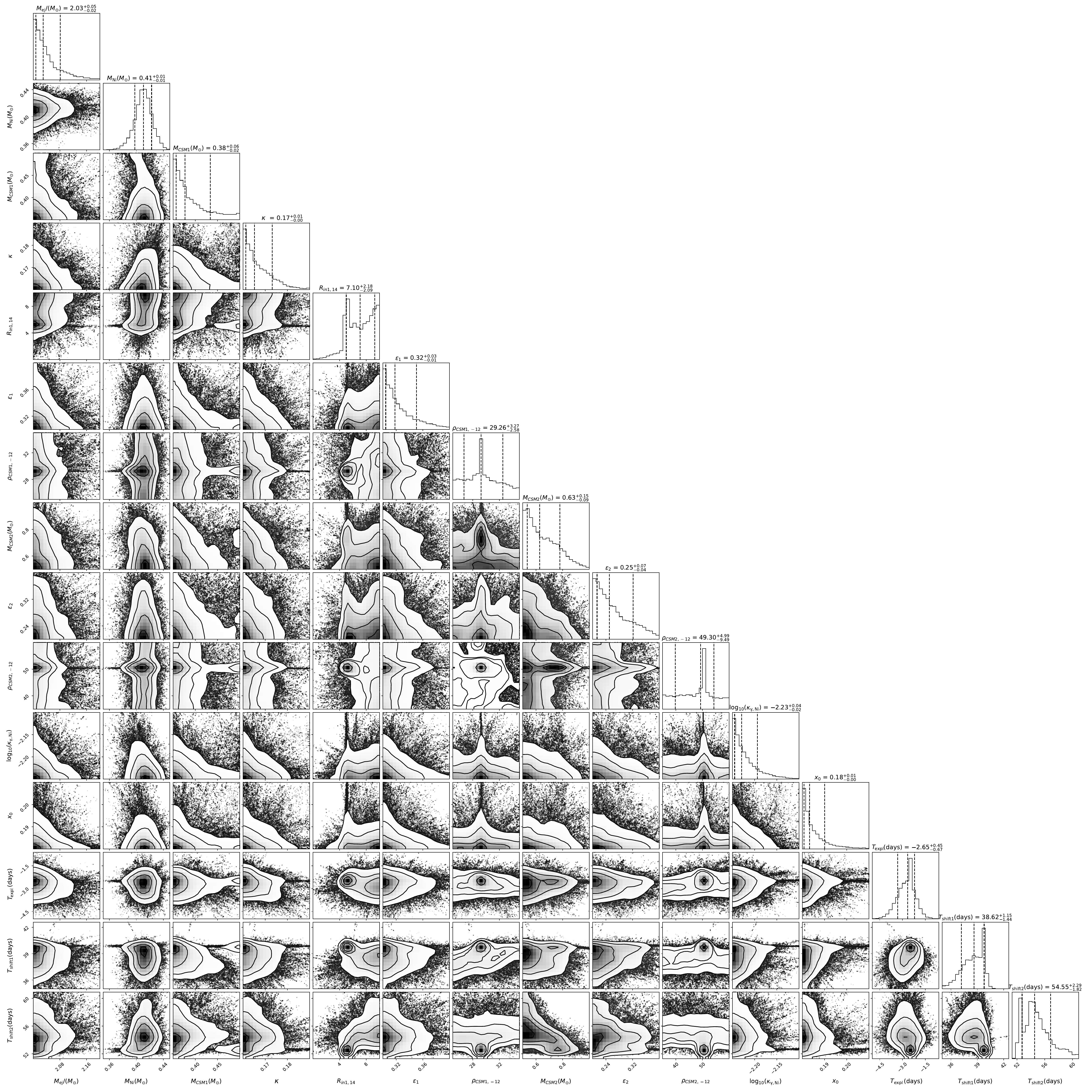}
    \caption{The corner plot for the CSI plus $^{56}$Ni model. The uncertainties are computed as the 16th and 84th percentiles (1$\sigma$) of the posterior samples along each axis.}
    \label{fig:CSI+56Ni_corner}
\end{figure*}

\subsection{The CSI Model} \label{the CSI model}

\citep{2012ApJ...746..121C} assumed a stationary photosphere within the CSM, due to its lower expansion velocity compared to SN ejecta, allowing the bolometric LC to be expressed as:

\begin{equation}
\begin{aligned}
&L_{\mathrm{CSI}}(t)=\frac{1}{t_0} e^{-\frac{t}{t_0}} \int_{t_{\mathrm{expl}}}^{t+t_{\mathrm{expl}}} e^{\frac{t^{\prime}}{t_0}} \times \\
&\left[L_{\mathrm{inp}, \mathrm{FS}}(t^{\prime}) + L_{\mathrm{inp}, \mathrm{RS}}(t^{\prime})\right] d t^{\prime},
\end{aligned}
\end{equation}
where $L_{\mathrm{inp}, \mathrm{FS}}(t^{\prime})$ and $L_{\mathrm{inp}, \mathrm{RS}}(t^{\prime})$ are the input luminosities from the forward shock (FS) and reverse shock (RS), $t_0=\frac{\kappa M_{\mathrm{CSM}, \mathrm{th}}}{\beta c R_{\mathrm{ph}}}$ is the diffusion timescale in the optically thick CSM \citep{2012ApJ...746..121C,2013ApJ...773...76C,2017ApJ...851...54W}. $M_{\mathrm{CSM}, \mathrm{th}}$ and $R_{\mathrm{ph}}$ are the mass and the photospheric radius of the optically thick CSM, respectively.

\par The CSI model has eight paramenters: the ejected mass $M_{\text{ej}}$ ($M_{\odot}$), the mass of the collided CSM $M_{\text{CSM}}$ ($M_{\odot}$), the optical opacity $\kappa$ (cm$^{2}$~g$^{-1}$), the density of the CSM $\rho_{\text{CSM}}$ (10$^{-12}$ g cm$^{-3}$), the initial radius of the collided CSM $R_{\text{CSM,in}}$ ($10^{14}$ cm), the conversion efficiency of the shock energy into radiation $\epsilon$, the dimensionless position parameter of break in the ejecta from the inner region to the outer region $x_{\text{0}}$, the trigger moment of the interactions $t_{\text{expl}}$ (days).
The fitting results of the CSI model are illustrated in Figure \ref{fig:CSI-single.pdf} and summarized in Table \ref{tab:CSM}.
And the corner plots of the CSI model are shown in Figure \ref{fig:CSI-corner.pdf}.

\subsection{The CSI Plus $^{56}$Ni Model} \label{the Multiple CSI plus 56NI}

\par Although single-component models are effective in capturing general trends in luminosity evolution, they are unable to reproduce multiple bumps in the declining light curve. Here we present a multiple CSI model to explain the complicated light curve of SN 2020tua. Such CSI model have been successfully applied to some SLSNe with bumpy features \citep{2018ApJ...856...59L,2020ApJ...891...98L,2023NatAs...7..779L}.
Considering the non-negligible contribution of the decay of $^{56}$Ni/Co, a comprehensive model should include both CSI and radioactive decay.
The hybrid CSI plus $^{56}$Ni model for output luminosity considers two distinct cases: a homogeneously expanding photosphere and a fixed photosphere. 
The expanding photosphere model is used for centrally located energy sources that heat the SN ejecta as it expands, whereas the fixed photosphere model is typically employed in CSI scenarios where the almost stationary CSM relative to the ejecta is heated. 
Thus the semi-analytic CSI plus $^{56}$Ni model can be written as \citep{2018ApJ...856...59L,2020ApJ...900..121L,2023NatAs...7..779L}\footnote{For hydrogen-rich SNe, such as type II-P SNe, which have a more abundant outer layer of hydrogen compared to other SNe, there is a need to consider a higher value for the optical opacity $\mathit{\kappa}$ \citep{2011MNRAS.415..199M,2018ApJ...865...95W,2012ApJ...746..121C}. That is because hydrogen-rich SNe originate from progenitor stars with massive hydrogen envelopes. The energy released from the recombination occurring in the extended hydrogen envelope of the progenitor manifests as a longer plateau feature in the light curve of hydrogen-rich SNe, rather than just small bumps.}:

\begin{equation}
\begin{aligned}
&L_{\mathrm{bolo}}(t)=\sum_{i}^{N} \frac{1}{t_{0,\mathit{i}}} e^{-\frac{t}{t_{0,\mathit{i}}}} \int_{0}^{t} e^{\frac{t^{\prime}}{t_{0,\mathit{i}}}} \\
& \times \left[L_{\mathrm{inp}, \mathrm{FS},\mathit{i}}(t^{\prime})+ L_{\mathrm{inp}, \mathrm{RS},\mathit{i}}(t^{\prime})\right] d t^{\prime}+\\
&\frac{2}{\tau_{m}}e^{-\frac{t^{2}}{\tau_{m}^{2}}} \int_0^t e^{\frac{t'^{2}}{\tau_{m}^{2}}} \frac{t'}{\tau_{m}}L_{\rm inp,Ni}(t')dt',
\end{aligned}
\end{equation}
where the diffusion timescale can be written as $t_{\text{0},\mathit{i}}=\sum\limits_{\mathit{j}=\mathit{i}}^{N} \frac{\mathit{\kappa}_{\mathit{j}}M_{\text{CSM,th},\mathit{j}}}{\beta cR_{\text{ph}}}$.
In this scenario, the main peak is mainly produced by the decay of $^{56}$Ni, while the bumps in the declining light curve is powered by the CSI. It is reasonable to assume that there are at least two collisions between the SN ejecta and the CSM.

\par There are several free parameters that need to be carefully considered: 
the gamma-ray opacity of \(^{56}\text{Ni}\) decay photons \(\kappa_{\gamma}\) (cm$^{2}$~g$^{-1}$),
the optical opacity $\kappa$ (cm$^{2}$~g$^{-1}$),
the $M_{\mathrm{ej,\mathit{i}}}$ ($M_{\odot}$) is the initial ejecta mass of the $i$th interaction with $M_{\mathrm{ej,2}}$= $M_{\mathrm{ej,1}}$ + $M_{\mathrm{CSM,1}}$, $M_{\mathrm{CSM,\mathit{i}}}$ is the mass of the $i$th collided CSM,
$R_{\mathrm{CSM,in,\mathit{i}}}$ is the initial radius of the $i$th collided CSM ($10^{14}$~cm) with $R_{\mathrm{CSM,in,2}} = R_{\mathrm{CSM,in,1}} + v_{\mathrm{SN}}~(t_{\mathrm{CSI,2}} - t_{\mathrm{CSI,1}})$, 
$\epsilon_{\mathit{i}}$ is the efficiency of the kinetic energy convert to radiation of the $i$th interaction,
$\rho_{\mathrm{CSM,in,\mathit{i}}}$ (10$^{-12}$ g cm$^{-3}$) is the density of the CSM at $R_{\mathrm{CSM,in,\mathit{i}}}$. 
the dimensionless position parameter of break in the ejecta from the inner region to the outer region $x_{\text{0}}$,
the epoch of the explosion $t_{\text{expl}}$ (days),
the time of the $i$th interaction between the ejecta and the $i$th CSM $t_{\mathrm{CSI,\mathit{i}}}$ (days).
\par The LC produced by the CSI plus $^{56}$Ni model are show in Figure \ref{fig:CSI+56Ni.pdf} and the best-fit paramenters are listed in Table \ref{tab:duoble+ni}.
And the corner plots of CSI plus $^{56}$Ni model are shown in Figure \ref{fig:CSI+56Ni_corner}.

\section{Discussion and Conclusion} \label{sec:discussion}
\subsection{The Properties of SN 2019tua} \label{sec:Comparing the Properties,The Properties of the SN2019tua}
\par Acrodding to the model that the CSI plus $^{56}$Ni model, the best-fitting parameters are $M_{\text{Ni}}=0.41_{-0.01}^{+0.01}~M_{\odot}$ and $M_{\text{ej}}=2.41~M_{\odot}$. 
The kinetic energy of SN 2019tua can be calculated, $\sim$ 0.5 $\times$ 10$^{51}$~erg. 
SN 2019tua has a lower kinetic energy compared to those of some SNe IIb: SN 2017gpn $\sim$ 3 $\times$ 10$^{51}$~erg \citep{2021MNRAS.501.5797B}; SN 2020acat $\sim$ 1.2 $\times$ 10$^{51}$~erg \citep{2022MNRAS.513.5540M}; SN 2011fu $\sim$ 1.3 $\times$ 10$^{51}$~erg \citep{2015MNRAS.454...95M}; ASASSN-14dq $\sim$ 1.24 $\times$ 10$^{51}$~erg \citep{2018MNRAS.480.2475S}; SN 2020cxd $\sim$ 4.3 $\times$ 10$^{51}$~erg \citep{2021A&A...655A..90Y}; SN 2003bg $\sim$ 5 $\times$ 10$^{51}$~erg \citep{2009ApJ...703.1612H}.
On the other hand, we can derive the mass-loss history of the event. 
Let \( \Delta t' \) denote the time interval between the ejection and the collision of the ejecta with the CSM, and let \( \Delta t = t_{\text{CSI}} - t_{\text{expl}} \) signify the interval between the SN explosion and the moment of shell collision. Thus, we have \( v_{\text{SN}}\Delta t = v_{\text{shell}} (\Delta t' + \Delta t) \), yielding \( \Delta t' = (v_{\text{SN}}/v_{\text{shell}} - 1)\Delta t \).
According to literature, the velocity of the progenitor shell for a type II SN ranges from 20 to 100 km s\(^{-1}\) \citep{2014ARA&A..52..487S}.
For SN 2019tua, with an observed velocity \( v_{\text{SN}} \) of approximately 5600 km s\(^{-1}\), and \( \Delta t_{\text{inner}} \) around 41.3 days, the derived time scale for the mass loss from the inner shell spans approximately 6.2 to 31.6 years (\( \sim 2271.5 - 11522.7 \) days). 
For \( \Delta t_{\text{outer}} \) approximately 57.2 days, the time scale for the mass loss from the outer shell is estimated to be 8.6 to 43.7 years (\( \sim 3146.0 - 15958.5 \) days).

\subsection{The Best Model} \label{sec:Comparing the model:Which is the Best Model?}
From a \( \chi^2 / \text{dof} \) standpoint, the \( ^{56}\text{Ni} \) model yields a satisfactory \( \chi^2 / \text{dof} = 0.37 \).
Magnetar models offer a superior fit to the data, evidenced by a \( \chi^2 / \text{dof} \) value of 0.98, although corroborative evidence for a magnetar-powered SN 2019tua is lacking \citep{2018MNRAS.475.2344V}.
Given the ongoing interactions between the SN ejecta and the CSM, especially near peak brightness, a single CSI model fits the data acceptably, attaining a \( \chi^2 / \text{dof} \) value of 0.35.
However, this model fails to accurately capture the secondary peak in the LC following the initial brightness peak.
Upon conducting AIC/BIC tests (with results detailed in accompanying tables), we find that the CSI plus \( ^{56}\text{Ni} \) model scores the highest, indicating its relative instability compared to other single models.
Contrastingly, the \( ^{56}\text{Ni} \) model most closely aligns with statistical criteria, displaying the lowest \( \chi^2 / \text{dof} \) and AIC/BIC values. 
In essence, no individual model can adequately account for the unique spectral features observed in SN 2019tua.

Additional spectral data offer further insights; narrow emission lines are discernible in the early spectra. 
And we have simply compared the redshift ($z$ = 0.010) of the host galaxy with the narrow emission lines in the spectrum SN 2019tua in the TNS\footnote{\url{https://www.wis-tns.org/object/2019tua}}, which are likely to originate from the host H II regions \citep{2017ApJ...850...89G}.
Moreover, there is unclear visible spectral feature of separation between H-shell and He-shell in Figure \ref{fig:spec1.pdf}, but there is obvious spectral feature of separation between H-shell and He-shell in Figure \ref{fig:spec2.pdf}.
The role of \( ^{56}\text{Ni} \) production during the SN explosion is non-negligible.
The presence of bumps in the LC is highly likely associated with interactions with the CSM. 
Taking into account the above situation, we adopted the CSI plus \( ^{56}\text{Ni} \) model, which also yielded a good \( \chi^2 / \text{dof} \) value of 0.85.
Although the \( \chi^2 / \text{dof} \) value may not fully elucidate these models due to uncertainties, the multi-component CSI plus \( ^{56}\text{Ni} \) model remains the most congruent with observed phenomena, and we also cannot rule out the possibility of the magnetar-powered model.

\par In this paper, we conducted comprehensive observations in the $\textit{BVRI}$~bands around the peak brightness of SN 2019tua and undertook extensive optical follow-up studies post-peak. Spectroscopically, the amalgamation of rising and near-peak spectra enabled a preliminary analysis of SN 2019tua. The combined photometric and essential spectroscopic data categorize SN 2019tua as a type IIb SN with robust observational evidence.

\par Data analysis reveals several key characteristics of the SN. Firstly, the spectrum of SN 2019tua displays a weak H$\alpha$ line and a narrow emission line, as well as a pronounced He absorption line, signifying the presence of CSM surrounding the progenitor. Given that the $L_{\text{peak}}$ of SN 2019tua exceeds that of certain type Ib SNe, a correlation with type Ib SNe is thereby confirmed. These observations collectively indicate that the progenitor of SN 2019tua had a low-mass, mostly H-stripped envelope, leaving only a thin H envelope and an exposed He layer \citep{2020A&A...641A.177M}. While the spectral features broadly align with other type IIb SNe, its bolometric LC appears to be comparatively brighter, necessitating further investigation.

\par SN 2019tua has a well-constrained explosion date. Unlike typical type IIb SNe, its LC lacks an early-time decline associated with a shock-cooling tail. SN 2019tua reached peak brightness approximately 25 days post-explosion, with a rate of brightness increase consistent with other type IIb SNe. Subsequently, a luminosity “bump” indicated interaction between the SN ejecta and CSM, suggestive of dense shells in the circumstellar environment. Various factors could contribute to such mass loss from the progenitor, and multiple shells resulting in a bumpy LC are conceivable.

\par After fitting the observed bolometric LC with three models, we conclude that SN 2019tua cannot be solely powered by the radioactive decay of $^{56}$Ni. Additionally, while no bursts of gamma-rays or X-rays were detected at the same location, we cannot rule out the possibility of a magnetar model.
A light curve primarily driven by radioactive decay of a moderate amount of $^{56}$Ni, augmented by CSM interaction, better replicates the observed behavior of SN 2019tua.
To account for the late LC bump, we employed a multiple-shell CSI model, with input parameters remaining within reasonable ranges.
\par In summary, the CSI model is substantiated by the presence of P$-$Cygni lines, suggesting that the progenitor of SN 2019tua with such features likely eject substantial amounts of material.
This finding is aligns with other observational results, such as luminosity fluctuations attributed to minor collisions.
Despite SN 2019tua has a relatively brighter luminosity compared to other type IIb SNe, SN 2019tua can be explained through CSI and radioactive decay, negating the necessity for additional energy sources.
The CSI plus \( ^{56}\text{Ni} \) model suggests that the progenitor of SN 2019tua had a pre-SN eruption. Multiple ejections before the explosion formed a multi-layered dense circumstellar shell around the SN. And the pre-SN eruption of SN 2019tua may produced by different mechanisms such as pulsational pair-instability supernova that undergo pulsations induced by pair production that may eject multiple interacting shells \citep{2007Natur.450..390W}, radiatively driven stellar winds that shed its envelope \citep{2007ARA&A..45..177C}, waves propagating through the star’s envelope may produce dense CSM in the few years leading up to core collapse \citep{2014ApJ...785...82S}. In the context of binary companion, the scenarios surrounding mass loss become considerably more intricate \citep{2020MNRAS.491.6000S}.

The distinctive bumps and high luminosity observed in SN 2019tua stand out when compared to other SNe of the same classification. These features are likely a result of significant mass loss from SN 2019tua's progenitor, which might have been more substantial than that experienced by progenitors of other type IIb SNe. This extensive mass loss likely led to the formation of multiple dense layers of CSM around SN 2019tua. The interaction between the ejecta and these denser layers of CSM, coupled with the ongoing energy contribution from the radioactive decay of $^{56}$Ni, are key to explaining SN 2019tua's higher luminosity and the presence of bumps in its light curve following the main peak. This scenario, supported by our $^{56}$Ni+CSM hybrid model, suggests that the unique interplay between enhanced mass loss, CSI, and nuclear decay significantly influences the observable characteristics of SN 2019tua, setting it apart from its type IIb counterparts.
Additionally, SN 2018ivc also powered by CSM interaction with SN ejecta, SN 2018ivc exhibits early cooling emission is a result of the SN shock breaking out of the stellar envelope, followed by multiple smaller bumps after its main peak \citep{2023ApJ...942...17M}. That suggests that the outer envelope mass mass of SN 2019tua is likely less than that of SN 2018ivc, resulting in SN 2019tua's inability to produce a early cooling emission but rather generating bumps after its main peak.

\section{Acknowledgements}
We are thankful to the referee very much for useful comments and constructive suggestions. 
This study is supported by the National Natural Science Foundation of China (grant Nos. 12373042, U1938201, 12133003, 12303050, 11973055 and 12173009), the Programme of Bagui Scholars Programme (WXG) and the Guangxi Science Foundation (grant Nos. 2020GXNSFDA238018). We acknowledge the support of the staff of the Xinglong 2.16m telescope. This work was partially supported by the Open Project Program of the Key Laboratory of Optical Astronomy, National Astronomical Observatories, Chinese Academy of Sciences.

\end{document}